\documentclass[prb,twocolumn,citeautoscript]{revtex4-1}

\usepackage{grffile}
\usepackage{graphicx}
\usepackage{amsmath}
\usepackage{amssymb}
\usepackage{color}
\usepackage{mathrsfs}
\usepackage{braket}
\usepackage{hyperref}

\usepackage{bm}\let\vec\bm 

\begin{document}

\title{Impurity coupled to a lattice with disorder}
\author{A.-M. Visuri}
\affiliation{Department of Quantum Matter Physics, University of Geneva, 24 quai Ernest-Ansermet, 1211 Geneva, Switzerland}

\author{C. Berthod}
\affiliation{Department of Quantum Matter Physics, University of Geneva, 24 quai Ernest-Ansermet, 1211 Geneva, Switzerland}

\author{T. Giamarchi}
\email{Thierry.Giamarchi@unige.ch}
\affiliation{Department of Quantum Matter Physics, University of Geneva, 24 quai Ernest-Ansermet, 1211 Geneva, Switzerland}

\begin{abstract}
We study the time-dependent occupation of an impurity state hybridized with a continuum of extended or localized states. Of particular interest is the return probability, which gives the long-time limit of the average impurity occupation. In the extended case, the return probability is zero unless there are bound states of the impurity and continuum. We present exact expressions for the return probability of an impurity state coupled to a lattice, and show that the existence of bound states depends on the dimension of the lattice. In a disordered lattice with localized eigenstates, the finite extent of the eigenstates results in a non-zero return probability. We investigate different parameter regimes numerically by exact diagonalization, and show that the return probability can serve as a measure of the localization length in the regime of weak hybridization and disorder. Possible experimental realizations with ultracold atoms are discussed.
\end{abstract}

\maketitle

\section{Introduction}

A discrete level coupled to a continuum of energies is a well-known problem in quantum optics \cite{Weisskopf1930, Cohen-Tannoudji1986, Mahan2000, Cohen-Tannoudji2004, Grynberg2010}. When the continuum is unbounded, the occupation of an initially occupied discrete level decays exponentially as the particle diffuses into the continuum. The decay law is not always exponential but depends on the density of states of the continuum. A particle in a discrete level could therefore be used as a probe of the continuum it is coupled to. For example, for a bounded continuum such as the energy band of a lattice, localized states outside the continuum can appear and lead to a nonzero occupation of the impurity level at infinite time. The limit of a zero-width continuum, on the other hand, corresponds to a two-state system with Rabi oscillations \cite{Cohen-Tannoudji2004}. A system with two localized (bound) states outside a finite continuum shows similar oscillations at long times, with an amplitude given by the overlap of the discrete level with the bound states. The discrete-level occupation and its long-time limit thus provide insight about the precise nature of the continuum.

The density of states, and therefore the decay law of a discrete level or impurity state, is modified in the case of a spatially disordered potential. A disordered system can exhibit the phenomenon of Anderson localization \cite{Anderson_absence1958} characterized by exponentially decaying wave functions \cite{Mott_impurity_conduction1961, Borland_disordered1963}. The localization results from interference between time-reversed scattering paths in a random medium and was first predicted for electrons in disordered crystals \cite{Anderson_absence1958}. In three-dimensional (3D) systems, localization occurs when the disorder potential exceeds a critical value, whereas in lower dimensions any nonzero disorder strength localizes the wave functions. In 3D, extended and localized states can coexist at different energies separated by so-called mobility edges, where the system's behavior changes from metallic to insulating.

Signs of Anderson localization have been observed in disordered systems as diverse as doped semiconductors \cite{Kastner_disordered_semiconductors1987}, light in random wave guides \cite{Chabanov_photon_localization2000, Chabanov_photon_localization2001, Chabanov_breakdown_of_diffusion2003, Shi_random_waveguides2014, Wiersma_localization_of_light1997, Storzer_localization_of_light2006, Schwartz_disordered_photonic2007, Lahini_photonic_lattices2009, Segev_localization_of_light2013}, and acoustic waves in mesoscopic glasses \cite{Aubry_recurrent_scattering2014, Cobus_mobility_gap2016}. Experiments with ultracold atoms have reported Anderson localization in one-dimensional (1D) random speckle potentials \cite{Billy_observation2008, Lugan_correlated_potentials2009}. Such potentials have a finite correlation length, which leads to an effective mobility edge even in 1D. A recent experiment demonstrated the existence of a single-particle mobility edge in a 1D potential formed by two incommensurate optical lattices \cite{Luschen_mobility_edge2017}. The Anderson metal-insulator transition in 3D has been investigated in speckle potentials \cite{Jendrzejewski_localization2012, McGehee_localization2013, Semeghini_mobility_edge2015} with somewhat inconclusive results \cite{Pasek_where2017}. 

Theoretically, disordered systems have been studied both with analytical and numerical tools \cite{Kramer_localization1993, Lagendijk_fifty_years2009}. In 1D, analytically solvable models exist, whereas in 2D and 3D, disordered systems have been treated by scaling theory \cite{Abrahams_scaling_theory1979}. For 1D systems, it is known that the localization length is of the order of the mean free path \cite{Berezinskii_kinetics1974, Abrikosov_conductivity1978}, the average distance between scattering events. In 2D, scaling theory predicts localization for any strength of disorder, but the localization length is an exponential function of the mean-free path and can be extremely large.

The localization length itself is a difficult quantity to measure, and localization is usually observed through the conductance of a material, or, in the case of ultracold atoms, the spatial distribution of the atoms. We focus in the present paper on using an impurity level to probe such systems. We consider a local observable, the probability that a particle initially in the impurity state returns to this state, and we investigate the relation between this observable and the localization length in the disordered lattice. We compare the cases of one- and two-dimensional lattices, for which localization properties are known to be different. The model studied here could be realized experimentally by coupling an impurity to an effectively 1D system, such as a quantum dot attached to a wire \cite{Kobayashi_Fano_resonance2004, Sato_quantum_dot2005} or using ultracold atoms in an optical potential with a local coupling to a different hyperfine state, as will be discussed below.

The plan of the paper is as follows: Sec.~\ref{sec:model_and_methods} introduces the model, the relevant quantities to characterize Anderson localization, a formal analytical solution for the return probability, and the numerical methods. Exact results for clean and strongly disordered systems are presented in Sec.~\ref{sec:analytic_results}. In Sec.~\ref{sec:numerical_results}, we check our numerics for the occupation probability of the impurity level against analytical expressions in the case of a non-disordered lattice, and we present numerical results for a disordered lattice. In Sec.~\ref{sec:experiments}, we discuss possible realizations of the model in experiments with ultracold atoms in optical potentials, and conclude in Sec.~\ref{sec:conclusions}.

\section{Model and methods}
\label{sec:model_and_methods}

\subsection{Return probability and parameter regimes}

\begin{figure}[tb]
\includegraphics[width=\linewidth]{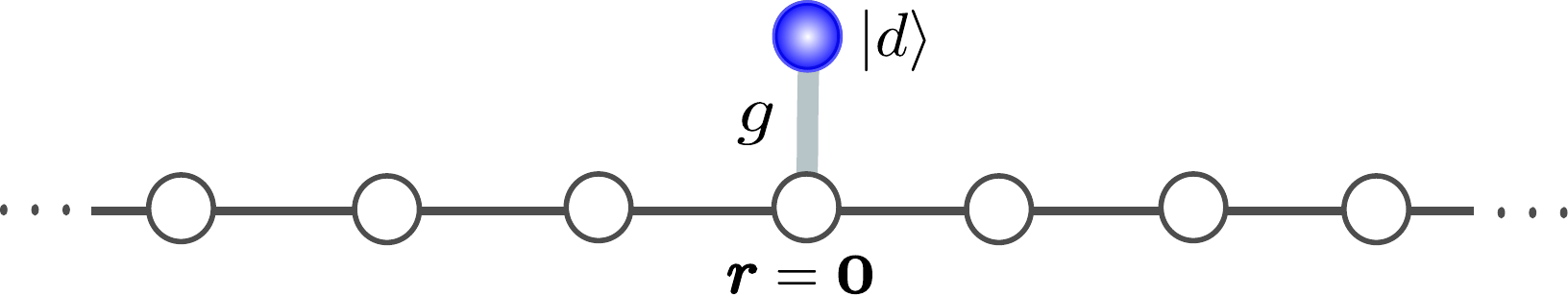}
\caption{The geometry of the model in the case of a one-dimensional lattice. An impurity state $\ket{d}$ is coupled with amplitude $g$ at site $\vec{r} = \vec{0}$.}
\label{fig:schematic}
\end{figure}

We consider an impurity state coupled to a disordered lattice. The Hamiltonian describing the system is
\begin{align}
\begin{split}
H &= H_0 + E_d n_d + g\big(c_d^\dagger c^{\phantom{\dagger}}_{\vec{r} = \vec{0}} + \text{H.c.}\big), \\
H_0 &= \sum_{\vec{k}} E_{\vec{k}} n_{\vec{k}} + \sum_{\vec{r}} V_{\vec{r}} n_{\vec{r}}.
\end{split}
\label{eq:Hamiltonian}
\end{align}
For convenience, we designate by $H_0$ the Hamiltonian of the disordered lattice without the impurity. Here, $\vec{k}$ and $\vec{r}$ are indices of momentum and position eigenstates in one, two, or three dimensions and $d$ denotes the impurity state. The number operators are defined as $n = c^\dagger c$, where $c^\dagger$ ($c$) is the creation (annihilation) operator. The discrete wave vectors $\vec{k}$ become continuous in the thermodynamic limit. $E_{\vec{k}}$ is the dispersion relation for particles moving on the clean (undisordered) lattice. We use the tight-binding form
\begin{equation}
E_{\vec{k}} = -2 J \sum_{i = 1}^D \cos(k_i),
\end{equation}
where $J$ is the hopping amplitude between nearest neighbor sites on a hypercubic lattice and $D$ is the dimensionality. We set the lattice spacing to one, choose units such that $\hbar\equiv 1$, and report energies relative to the half-bandwidth $W = 2 J D$. In Eq.~(\ref{eq:Hamiltonian}), $V_{\vec{r}}$ denotes a random uncorrelated on-site disorder that is uniformly distributed between $-V$ and $+V$. The energy at the impurity site is $E_d$ and the coupling between the impurity state and the site $\vec{r} = \vec{0}$ of the lattice is denoted by $g$. The geometry of the model is illustrated in Fig.~\ref{fig:schematic}.

We introduce the set of single-particle eigenstates $|\alpha\rangle$ of $H_0$ with eigenvalues $E_{\alpha}$. In a disordered system, the eigenstates can be exponentially localized as $|\psi_{\alpha}(\vec{r})| \sim e^{-|\vec{r}-\vec{r}_{\alpha}|/\xi}$, where $\xi$ is the localization length. In three dimensions, localization occurs above a critical value of the disorder strength, whereas a weakly disordered system is conducting with extended eigenstates. In one and two dimensions, the eigenstates are localized for any nonzero disorder strength. However, in 2D the localization length can be extremely large as it depends exponentially on the mean-free path $\ell$ \cite{Lee_disordered1985},
\begin{equation*}
\xi \sim \ell\, e^{\frac{\pi}{2}k_{\mathrm{F}}\ell},
\end{equation*}
where $k_{\mathrm{F}}$ is the Fermi wave vector. Localization is strongest in 1D where the localization length is twice the mean-free path \cite{Thouless_localization_distance1973}. As the strength of the disorder potential varies between the clean limit with $V=0$ where all eigenstates are extended to the limit $V \gg W$ where all states are strongly localized, both $\ell\sim V^{-2}$ and $\xi$ decrease from infinity to lengths comparable with the lattice spacing. The question we address is whether the impurity state could serve as a probe of localization, specifically, whether the non-equilibrium population of the impurity state measures the localization length in the lattice.

We prepare the system at time $t=0$ with one particle occupying the impurity state $\ket{d}$ and no particle in the lattice and measure the probability of finding the particle in the state $\ket{d}$ as a function of time. The quantity of interest is the return probability, or survival probability, of the impurity state -- the infinite-time limit of the time-averaged
occupation
\begin{equation}
Q_d = \lim_{T \rightarrow \infty}\frac{1}{T}\int_0^T \overline{|\langle d | e^{-i H t} | d \rangle|^2}.
\label{eq:return_probability}
\end{equation}
The bar on the right-hand side represents the disorder average, which is done in the case of a disordered lattice. For $V = 0$, the problem reduces to the textbook problem of a discrete level coupled to a smooth continuum, which was solved early on at leading order in the coupling and for an unbounded continuum \cite{Weisskopf1930}. The result is an exponential decay of the discrete-level occupation with a decay rate given by the Fermi golden rule \cite{Weisskopf1930, Cohen-Tannoudji1986, Cohen-Tannoudji2004, Grynberg2010}. In the opposite limit $g \gg W$, the problem approaches a two-level system with Rabi oscillations. The case of a finite-width continuum, either bounded from below, above, or both, is more intriguing because a finite occupation of the discrete level can survive even in the limit $t \rightarrow \infty$ \cite{Economou1979, Kogan2007}. This is due to the emergence of impurity-induced bound states outside the continuum. Two such bound states lead to Rabi-like oscillations and one single bound state to a constant occupation at long times. In Sec.~\ref{sec:formal_solution}, we discuss a formal solution that is exact at all orders in $g$ for $V = 0$ and a continuum of finite bandwidth. In particular, we show how the dimensionality of the lattice is connected with the existence of either zero, one, or two bound states. In Sec.~\ref{sec:numerical_results_no_disorder}, we crosscheck this solution numerically in dimensions $D=1,2,3$ by means of exact diagonalization and expansion of the evolution operator on Chebyshev polynomials.

For $V>0$, one can distinguish various regimes. In the clean limit $0<V\ll W$, the physics is similar to that for $V=0$ with small perturbative corrections in $V/W$.  One exception is the weak-coupling region $g<V$, where the corrections are large as will be seen in Sec.~\ref{sec:numerical_results}. If $V\gg W$, the eigenstates $\ket{\alpha}$ are strongly localized and eventually confined to a single site. In this limit, the impurity state is effectively coupled to only one lattice site resulting in Rabi oscillations for any coupling $g$ (Sec.~\ref{sec:strong_disorder}). The most interesting regime---regarding the information that the return probability may hold about localization---is $V\lesssim W$ and $g<V$, where the localization length varies and the coupling $g$ is not strongly perturbing the lattice.

\subsection{Measures of Anderson localization}
\label{sec:localization_measures}

There are several ways of measuring whether a system is localized. We briefly discuss three of them here, namely the \emph{lattice} return probability, the inverse participation ratio, and the transport localization length. The lattice return probability to a site $\ket{\vec{r}}$ is analogous to the return probability of Eq.~(\ref{eq:return_probability}) and has a simple expression in terms of the eigenstates $\ket{\alpha}$ of $H_0$ (see Appendix~\ref{app:return_probability}):
\begin{equation}
Q_{\vec{r}} = \lim_{T \rightarrow \infty}\frac{1}{T}\int_0^T \overline{|\langle\vec{r} | e^{-i H_0 t} | \vec{r}\rangle|^2}
= \sum_{\alpha}\overline{|\langle\vec{r}|\alpha\rangle|^4}.
\label{eq:Qr_eigenstates}
\end{equation}
The bar denotes disorder averaging as in Eq.~(\ref{eq:return_probability}). The lattice return probability reaches the maximum value of one when the eigenstates are maximally localized, $\ket{\alpha} = \ket{\vec{r}_{\alpha}}$, and a minimum value of $1/N$ for extended plane-wave states, where $N$ is the number of lattice sites.

The inverse participation ratio provides a measure of the localized character of a given state:
\begin{equation}
\text{IPR}_{\alpha} = \sum_{\vec{r}}|\langle\vec{r}|\alpha\rangle|^4.
\end{equation}
Like $Q_{\vec{r}}$, the inverse participation ratio increases from $1/N$ to 1 as the state becomes more and more localized. Without the disorder-average in Eq.~(\ref{eq:Qr_eigenstates}), $\sum_{\vec{r}} Q_{\vec{r}} = \sum_{\alpha} \text{IPR}_{\alpha}$, such that the average values of $Q_{\vec{r}}$ and $\text{IPR}_{\alpha}$ are identical over the interval $[1/N,1]$. A numerical study furthermore showed that these values have similar distributions \cite{Luck2016}. It is convenient to represent the IPR of a state by an equivalent length defined as the characteristic length of an exponentially-localized wave function in the continuum with the same IPR value. By calculating the IPR for a state $\psi(\vec{r}) \sim e^{-r/\xi}$ in dimension $D$, we deduce the expression of the equivalent length as
\begin{equation}\label{eq:xi_alpha}
	\xi_{\alpha}^{-1}=2\left[\pi^{\frac{D-1}{2}}\Gamma\left(\textstyle\frac{D+1}{2}\right)
	\mathrm{IPR}_{\alpha}\right]^{1/D},
\end{equation}
where $\Gamma$ is the Euler gamma function. We consider the disorder-averaged quantities $\xi$ and $\xi^{-1}$, which we obtain numerically as functions of energy by a binning procedure. The values of $\xi_{\alpha}$ and $\xi_{\alpha}^{-1}$ calculated from $\text{IPR}_{\alpha}$ are binned according to the corresponding eigenenergy $E_{\alpha}$ and averaged in each bin, and these values are averaged over the different disorder realizations.

The transport localization length characterizes the exponential decrease of the ballistic conductance in a disordered conductor of increasing length. The conductance can be for instance related to the Green's function, which gives in 1D the following expression for the localization length:
\begin{equation}
\frac{1}{\lambda(E)} = -\lim_{L \rightarrow \infty} \frac{1}{2L} \overline{\ln \frac{|G(0, L, E)|^2}{|G(0, 0, E)|^2}}.
\label{eq:conductance}
\end{equation}
The quantity $G(r,r',E)=\braket{r|(E+i0-H_0)^{-1}|r'}$ is the retarded Green's function for a disordered chain of length $L$ connected with two ideal leads. The symbol $i0$ denotes an infinitesimal imaginary part. In higher dimensions, we must sum all conduction channels and replace $|G(0,L,E)|^2$ by $\sum_{\vec{r}_0\vec{r}_L}|G(\vec{r}_0,\vec{r}_L,E)|^2$, where $\vec{r}_0$ and $\vec{r}_L$ represent all sites in contact with the left and right lead, respectively. Likewise, the normalization $|G(0,0,E)|^2$ is replaced by $\sum_{\vec{r}_0\vec{r}_0'}|G(\vec{r}_0,\vec{r}_0',E)|^2$.

The various measures of localization give qualitatively consistent although generally different results. The first two measures are ideal when exact diagonalization is possible and they converge provided that the linear system size is larger than $\xi_{\alpha}$ for all $\alpha$. The third one is convenient in 2D and 3D when $\xi_{\alpha}^D$ exceeds system sizes attainable by exact diagonalization, thanks to efficient algorithms for computing the Green's function or the transmission coefficients \cite{MacKinnon-1981, Pichard-1981}. The convergence of the transport localization length with system size is slow, though, such that a finite-size scaling analysis is required in order to extract reliable values in the thermodynamic-limit  \cite{Mackinnon_scaling_theory1983}.

\subsection{Formal solution for the return probability}
\label{sec:formal_solution}

The time evolution entering Eq.~(\ref{eq:return_probability}) admits a closed form than involves the self-energy of the impurity state $|d\rangle$ \cite{Economou1979, Mahan2000}. This can be shown for instance by means of a Laplace transform as done in Appendix~\ref{app:analytic_Laplace} or by using the equation of motion of the impurity Green's function as done in  Appendix~\ref{app:analytic_Greens_function}. The impurity self-energy accounts for the hybridization of the level with the lattice. When the impurity is coupled to a single site $\vec{r}=\vec{0}$ like in Eq.~(\ref{eq:Hamiltonian}), the self-energy is simply proportional to the local lattice Green's function at that site:
	\begin{align}\label{eq:self-energy}
		\nonumber
		\Sigma(E)&\equiv\Sigma_1(E)+i\Sigma_2(E)\\
		&=g^2G(\vec{0},\vec{0},E)=g^2\sum_{\alpha}\frac{|\langle\vec{0}|\alpha\rangle|^2}{E-E_{\alpha}+i0}.
	\end{align}
The factor $g^2$ accounts for the particle jumping in and out of the lattice and the propagator $G(\vec{0},\vec{0},E)=\braket{\vec{0}|(E+i0-H_0)^{-1}|\vec{0}}$ represents the excursion of the particle in the lattice from site $\vec{0}$ and back to site $\vec{0}$. The self-energy determines the spectral function of the impurity, 
\begin{equation}
A(E) = -\frac{1}{\pi} \text{Im} \left[ \frac{1}{E - E_d - \Sigma(E) + i0} \right],
\label{eq:impurity_spectral_function}
\end{equation}
which approaches a delta function at energy $E_d$ as the coupling $g$ approaches zero. The time-dependent amplitude on the impurity is related to these quantities as follows:
	\begin{multline}
		\braket{d|e^{-iHt}|d} = \int_{E_{\min}}^{E_{\max}}dE\,e^{-iEt}A(E)\\
		+\sum_{E_b}\frac{e^{-iE_bt}}{1-\partial_E \Sigma_1(E)|_{E=E_b}}.
		\label{eq:wavefunction_amplitude}
	\end{multline}
We split the amplitude in two terms in order to emphasize the role of the impurity-induced bound states outside the continuum, when they exist. The continuum bounded by $E_{\min}$ and $E_{\max}$ is defined by the condition $\Sigma_2(E)\neq0$: as seen in Eq.~(\ref{eq:self-energy}), this covers the spectral range of the lattice energies $E_{\alpha}$, which because of disorder extends beyond the bandwidth of the clean system. Outside the continuum, Eq.~(\ref{eq:impurity_spectral_function}) shows that the spectral function becomes $A(E)=\delta\left[E-E_d-\Sigma_1(E)\right]$. Therefore, if bound states exist outside the continuum, they are the solutions $E_b$ of
\begin{equation}
E_b - E_d - \Sigma_1(E_b) = 0,\qquad \Sigma_2(E_b) = 0.
\label{eq:bound_state_criterion}
\end{equation}
Their contribution to the impurity population is the second term on the right-hand side of Eq.~(\ref{eq:wavefunction_amplitude}), which, given the above form of the spectral function, could also be accounted for by extending the integration limits in the first term to $\pm\infty$. The interest of separating the two terms appears when considering Eq.~(\ref{eq:return_probability}): the contribution of a smooth continuum vanishes at long times as a power law controlled by the continuum boundaries,\footnote{The contribution of the continuum to $\braket{d|e^{-iHt}|d}$ vanishes as $1/t^{1+\nu}$, where $\nu$ is the smallest of the two exponents characterizing the continuum boundaries, for instance $A(E)\propto|E-E_{\max}|^{\nu}$.} such that the long-time occupation is set by the second term in Eq.~(\ref{eq:wavefunction_amplitude}). Note that the stationary Schr{\"o}dinger equation for the bound states $\ket{\psi_b}$ gives Eq.~(\ref{eq:bound_state_criterion}) as the eigenvalue equation and $|\langle d|\psi_b\rangle|^2=1/[1-\partial_E \Sigma_1(E)|_{E=E_b}]$. Hence the \emph{amplitude} entering Eq.~(\ref{eq:wavefunction_amplitude}) is equal to the \emph{probability} for the particle to be in the bound state at time zero.

\begin{figure}[tb]
\includegraphics[width=0.7\linewidth]{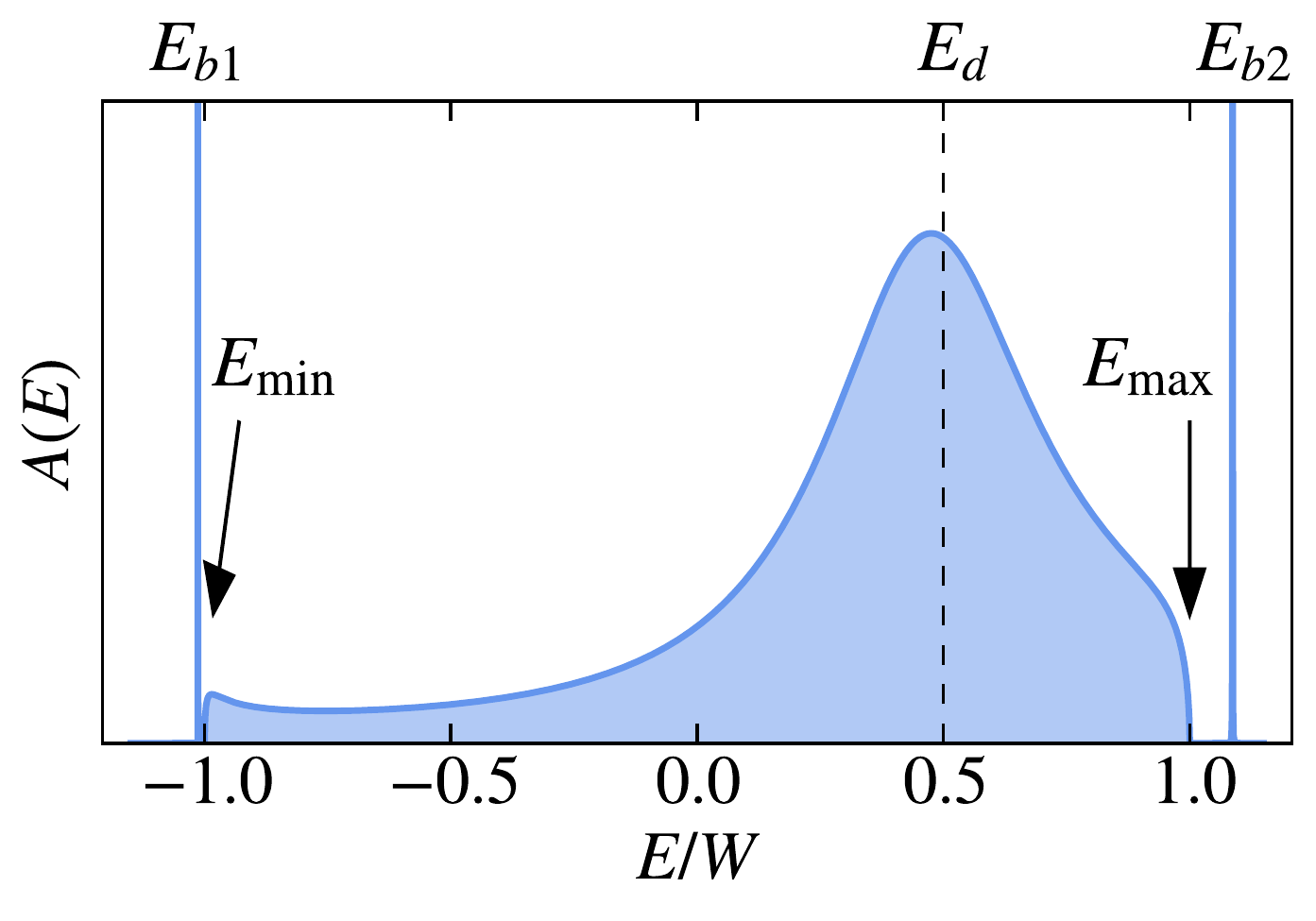}
\caption{Impurity spectral function for a clean 1D lattice and an impurity-lattice coupling $g=0.5\,W$. The initial delta function at the energy $E_d$ of the impurity gets shifted and broadened over the lattice continuum, which extends from $E_{\min}$ to $E_{\max}$; this explains the exponential decay of the impurity-level occupation at short time. The edges of the continuum at $E_{\min}$ and $E_{\max}$ control the power-law decay at intermediate time. Bound states of the impurity and lattice furthermore emerge below and above the continuum, which explains the saturation of the occupation to a finite value at long time.}
\label{fig:spectral_function}
\end{figure}

For a clean system in the thermodynamic limit, $E_{\min}$ and $E_{\max}$ coincide with the band edges of the lattice and $A(E)$ is continuous between these limits. Depending upon the dimensionality and impurity-lattice coupling, Eq.~(\ref{eq:bound_state_criterion}) can have zero, one, or two solutions with the corresponding wave functions centered at the impurity site and decaying to zero away from it, as discussed in Sec.~\ref{sec:no_disorder}. Figure~\ref{fig:spectral_function} shows the continuum and the bound states for a clean 1D lattice.

When disorder is present and weak, $E_{\min}$ and $E_{\max}$ move below and above the lattice band edges by an amount of order $V$. If this exceeds the energy of the bound states, the latter disappear. At the same time, the continuum $A(E)$ becomes itself discontinuous and gives a finite contribution to the long-time impurity occupation. As the disorder gets stronger, localization implies that the impurity is coupled to a \emph{finite} number of states in the lattice, even in the thermodynamic limit: the continuum $A(E)$ transforms into a finite set of delta peaks and the resulting long-time occupation is periodic.

Finally, for a clean or disordered system of finite size the spectral function $A(E)$ is discrete and each level contributes to the impurity occupation a term like the second term of Eq.~(\ref{eq:wavefunction_amplitude}). The quantity $\partial_E \Sigma_1(E)$ is in principle well-defined because, while $\Sigma_2(E)$ is made of Dirac delta functions at the energies $E_{\alpha}$, $\Sigma_1(E)$ is continuous in-between consecutive values of $E_{\alpha}$. If not for accidental degeneracies, the discrete levels of $H$ are different from those of $H_0$ and fall in regions where $\partial_E \Sigma_1(E)$ exists.

\subsection{Numerical methods}

In 1D, we calculate the occupation probability of the impurity state numerically by exact diagonalization. It allows us to reach sufficiently large system sizes $L$ compared to the localization length $\xi$ so that the results do not depend on $L$. In 2D, we use exact diagonalization where applicable. Since the size of systems solvable by exact diagonalization is limited, we use an expansion on Chebyshev polynomials for large 2D lattices and in 3D. The Chebyshev expansion of the time evolution operator is written as (see Appendix~\ref{app:chebyshev})
\begin{equation}
e^{-i H t} \approx e^{-i b t} \sum_{m = 0}^M (2 - \delta_{m 0}) (-i)^m J_m(a t) T_m(\tilde{H}),
\label{eq:chebyshev_expansion}
\end{equation}
where $T_m(x) = \cos(m \arccos x)$ are the Chebyshev polynomials defined for $x \in [-1, 1]$ and $J_m(x)$ is the Bessel function. The argument $\tilde{H}$ is the scaled Hamiltonian $\tilde{H} = (H - b)/a$, where $b$ is the middle and $2a$ the width of the spectrum of $H$ (or an upper bound on it). The scaled Hamiltonian is therefore dimensionless and has eigenvalues in the range $[-1, 1]$. The expansion (\ref{eq:chebyshev_expansion}) is exact for $M=\infty$ and truncated to order $M$ for calculations. It is analytic in $t$ and valid up to a time $t_M\approx M/a$. The order $M$ is chosen such that the return probability $Q_d$ calculated for $T=t_M$ is converged. We also use the Chebyshev expansion for evaluating the lattice Green's function thanks to the expansion (Appendix~\ref{app:chebyshev}),
\begin{multline}\label{eq:chebyshev_expansionG}
(E + i0 - H_0)^{-1}\approx\\ \frac{1}{a} \sum_{m = 0}^M K_m^M \frac{i (\delta_{m0} - 2) e^{-im\arccos (\tilde{E})}}{\sqrt{1 - \tilde{E}^2}} T_m(\tilde{H}_0).
\end{multline}
$\tilde{E}=(E-b)/a$ is the energy rescaled like the Hamiltonian. Again, the expansion is exact for $M=\infty$ with $K_m^M=1$. When it is truncated to order $M$, Gibbs oscillations develop \cite{Weisse_kernel2006}, which are suppressed by the kernel $K_m^M$. We use the Fej{\'e}r kernel $K_m^M = (1 - m/M)$ and check convergence with respect to $M$.

\section{Analytical results}
\label{sec:analytic_results}

It is know that the population of a discrete level coupled to a smooth continuum of infinite width decays exponentially with time at a rate given by the Fermi golden rule \cite{Weisskopf1930, Cohen-Tannoudji1986, Cohen-Tannoudji2004, Grynberg2010}. The same result can be obtained as a short-memory approximation.\footnote{The short-memory approximation amounts to replacing $\protect\langle\vec{0}|\alpha\protect\rangle$ by a constant in Eq.~(\ref{eq:self-energy_d}) and assuming a flat spectral density, which gives $M(t)=2\pi g^2N(E_d)\delta(t)$ with $N(E)$ the density of states of the lattice.} When the continuum has finite width, the time evolution is modified by the bound states leading to a finite return probability \cite{Kogan2007}.

\subsection{Clean lattice without disorder}
\label{sec:no_disorder}

\begin{figure}[tb]
\includegraphics[width=0.7\linewidth]{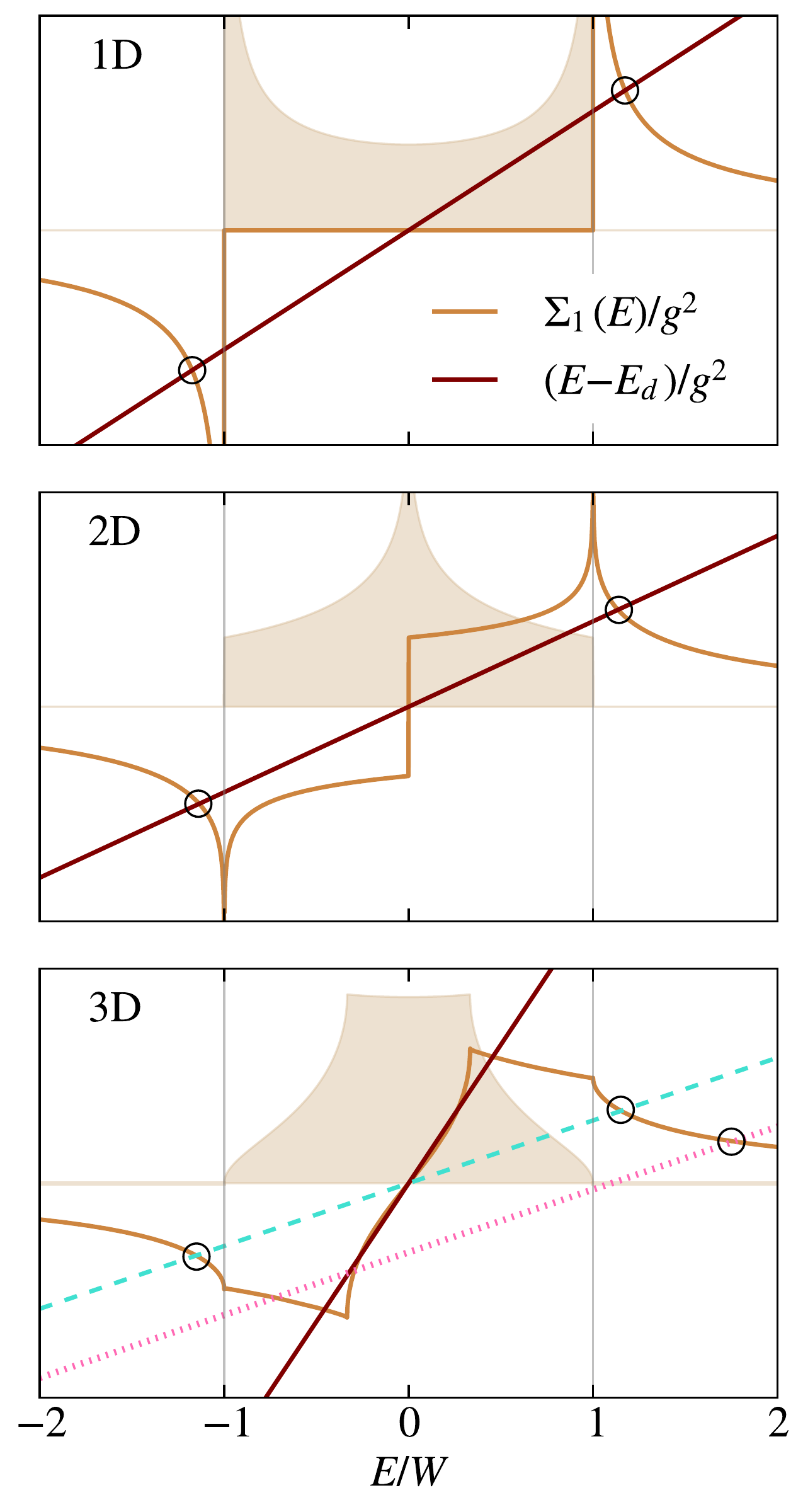}
\caption{Graphical solution of Eq.~(\ref{eq:bound_state_criterion}) for the bound states in dimensions 1, 2, and 3. The shaded curves show $-\Sigma_2(E)/g^2$, which is proportional to the lattice density of states and defines the energy range where bound states are forbidden. The band edges are marked with gray vertical lines. The brown solid lines show $\Sigma_1(E)/g^2$ and the dark solid lines show $(E-E_d)/g^2$ for $E_d = 0$. The intersections outside the forbidden range correspond to bound states and are marked with circles. For the 1D and 2D lattices, $\Sigma_1(E)$ diverges at the band edges and therefore there are always two bound states. For the 3D lattice, the number of intersections can be 0 (solid line), 2 (dashed line), or 1 (dotted line). The solid and dashed lines correspond to $E_d = 0$ while the dotted line has $E_d > 0$.}
\label{fig:intersections}
\end{figure}

As already pointed out, in the thermodynamic limit the first term in Eq.~(\ref{eq:wavefunction_amplitude}) vanishes for $t \rightarrow \infty$. The infinite-time limit of Eq.~(\ref{eq:return_probability}) is therefore given by the second term as
\begin{equation}
Q_d = \sum_{E_b} \frac{1}{\left| 1 - \partial_E \Sigma_1(E)|_{E = E_b} \right|^2} \: .
\label{eq:return_probability_exact}
\end{equation}
The bound-state energies solutions of Eq.~(\ref{eq:bound_state_criterion}) can be found graphically as the intersections of $\Sigma_1(E)$ and $E - E_d$ where $\Sigma_2(E)=0$, that is, outside the band. Equation~(\ref{eq:self-energy}) indeed shows that $\Sigma_2(E)=-\pi g^2N(E)$ is simply proportional to the lattice density of states $N(E)$ in the clean system. Since the real part of the self-energy follows different power laws at the band edges in different dimensions, the existence of bound states depends on the dimension of the lattice. Figure~\ref{fig:intersections} illustrates how the number of intersections depends on the dimension and the parameters $g$ and $E_d$. The figure shows the line $(E - E_d)/g^2$ for various values of $g$ and $E_d$ as well as $\Sigma_1(E)/g^2$ and $-\Sigma_2(E)/g^2$. The last two quantities are independent of $g$ and the latter highlights the range of the continuum where bound states cannot exist. In 1D and 2D, $\Sigma_1(E)$ has square-root and logarithmic singularities at the band edges, respectively. Therefore, there are two intersections for any values of $E_d$ and $g\neq 0$. For a 3D or higher-dimensional lattice, $\Sigma_1(E)$ is finite at the band edges. Therefore, a critical coupling $g_c$ exists below which there is no bound state. For $E_d=0$ and $g>g_c$, two symmetric bound states form like in 1D and 2D. If $E_d\neq 0$, we can have a situation where only one bound state exists, either above the band if $E_d>0$ or below the band if $E_d<0$.

The bound-state energies are exactly known in 1D, while one must resort to numerics in 2D and 3D. In 1D, we have $\Sigma_1(E)=g^2/(2\pi)\int_{-\pi}^{\pi}dk/(E+W\cos k)=(g^2/W)\mathrm{sign}(x)\mathrm{Re}\left[(x^2-1)^{-1/2}\right]$ with $x=E/W$. The equation giving the bound-state energy at positive $x$ becomes $(g/W)^2=(x-y)\sqrt{x^2-1}$ with $y=E_d/W$. The general solution is complicated but simplifies for $E_d=0$ to
\begin{equation*}
\frac{E_b}{W} = \pm\sqrt{\textstyle\frac{1}{2} + \frac{1}{2}\sqrt{1 + 4(g/W)^4}}.
\end{equation*}
Inserting these expressions in Eq.~(\ref{eq:return_probability_exact}) yields
\begin{equation}
Q_d =
    \begin{cases}
      1 & g=0 \\
      \frac{\left[ \sqrt{1 + 4(g/W)^4} - 1 \right]^2}{2 \left[1 + 4 (g/W)^4 \right]} & g \neq 0.
    \end{cases}
\label{eq:Qd1D}
\end{equation}
The return probability increases very slowly like $2(g/W)^8$ at small $g$ and approaches $1/2$ from below with a correction $-\frac{1}{2}(W/g)^2$ at large $g$.

\subsection{Strong impurity-lattice coupling}
\label{sec:strong_g}

For large $g$, the problem can be formulated as an effective two-level Hamiltonian of the form
	\begin{equation*}
		H_{\mathrm{eff}}=\begin{pmatrix}E_d & g \\ g & V_{\vec{r}=\vec{0}}+\Delta(E)\end{pmatrix},
	\end{equation*}
where the function $\Delta(E)=J^2\sum_{\vec{s}\vec{s}'}\bar{G}(\vec{s},\vec{s}',E)$ takes into account the hybridization of the site $\vec{r}=\vec{0}$ with the rest of the lattice. The sum runs over all sites $\vec{s}$ connected with the site $\vec{0}$ and $\bar{G}$ is the ``cavity'' Green's function, i.e., the Green's function of the lattice without the site $\vec{0}$. For $g\gg V,W$, the eigenvalues of $H_{\mathrm{eff}}$ approach $\pm g$ and Eq.~(\ref{eq:self-energy}) shows that $-\partial_E\Sigma_1(E)_{E=\pm g}$ approaches $\sum_{\alpha}|\langle\vec{0}|\alpha\rangle|^2=1$. According to Eqs~(\ref{eq:wavefunction_amplitude}) and (\ref{eq:return_probability}), the limiting value for $g\to\infty$ is therefore $Q_d=1/2$. If $E_d, W\ll V\ll g$, the leading correction to the eigenvalues is $V_{\vec{0}}/2\pm V_{\vec{0}}^2/(8g)$ and the value of $-\partial_E\Sigma_1(E)$ becomes $g^2/[\pm g+V_{\vec{0}}/2\pm V_{\vec{0}}^2/(8g)]^2$. Evaluating $Q_d$ with this expression and performing the impurity average, we get the asymptotic behavior
	\begin{equation}\label{eq:Qd_largeV}
		Q_d=\frac{1}{2}+\frac{2}{3}\left(\frac{V}{4g}\right)^2\qquad (E_d, W\ll V\ll g),
	\end{equation}
which shows that $Q_d$ approaches $1/2$ from above in this regime. In the weak-disorder regime, $E_d, V\ll W\ll g$, the leading correction to the eigenvalues is $\Delta(E)/2$ and we can approximate $\Delta(E)$ by $2DJ^2/E$. The reason is that there are $2D$ sites $\vec{s}$ connected with the site $\vec{0}$ and the high-energy limit of $\bar{G}(\vec{s},\vec{s},E)$ is $1/E$, while $\bar{G}(\vec{s},\vec{s}',E)$ is of order $1/E^2$ for $\vec{s}\neq\vec{s}'$. Solving for the eigenvalues and expanding the resulting $Q_d$ in powers of $W/g$ yields a behavior consistent with the one we deduced from Eq.~(\ref{eq:Qd1D}),
	\begin{equation}\label{eq:Qd_clean_high_g}
		Q_d=\frac{1}{2}-\frac{1}{2D}\left(\frac{W}{g}\right)^2\qquad (E_d, V\ll W\ll g),
	\end{equation}
which shows that for weak disorder the asymptotic value of $1/2$ is approached from below. Note that the consistency of the $W/g$ expansion requires us to include the sub-leading term in the high-energy expansion of the self-energy, namely $-\partial_E\Sigma_1(E)=g^2/E^2+3g^2W^2/(2DE^4)$. Our numerical data confirm these asymptotic results.

\subsection{Model for strong disorder}
\label{sec:strong_disorder}

When the strength of the disorder exceeds the bandwidth, the eigenstates $\ket{\alpha}$ are strongly localized and eventually confined to a single site in the limit $W/V\to0$. One of these states sits on the site $\vec{r}=\vec{0}$ and forms a two-level subsystem with the impurity, while all other lattice eigenstates are decoupled. The energy of the state localized at site $\vec{0}$ takes arbitrary values in the range $[-(W+V),W+V]$ as the disorder configurations are scanned. We show in Appendix~\ref{app:strong_disorder} that if all values in this interval are equally likely, the disorder-averaged return probability becomes in this limit, for $E_d=0$:
\begin{equation}
Q_d = 1 - \frac{g}{W + V} \arctan \left( \frac{W + V}{2 g} \right).
\label{eq:strong_disorder}
\end{equation}
As a function of $g$, this decreases linearly at small $g$ like $1-g(\pi/2)/(W+V)$ and approaches $1/2$ from above at large $g$ with the asymptotic correction $\frac{2}{3}[(W+V)/(4g)]^2$, consistently with Eq.~(\ref{eq:Qd_largeV}). At sufficiently strong disorder, the particle gets locked on the impurity and $Q_d$ approaches unity.

\section{Numerical results}
\label{sec:numerical_results}

\subsection{No disorder}
\label{sec:numerical_results_no_disorder}

\begin{figure}[b]
\includegraphics[width=\linewidth]{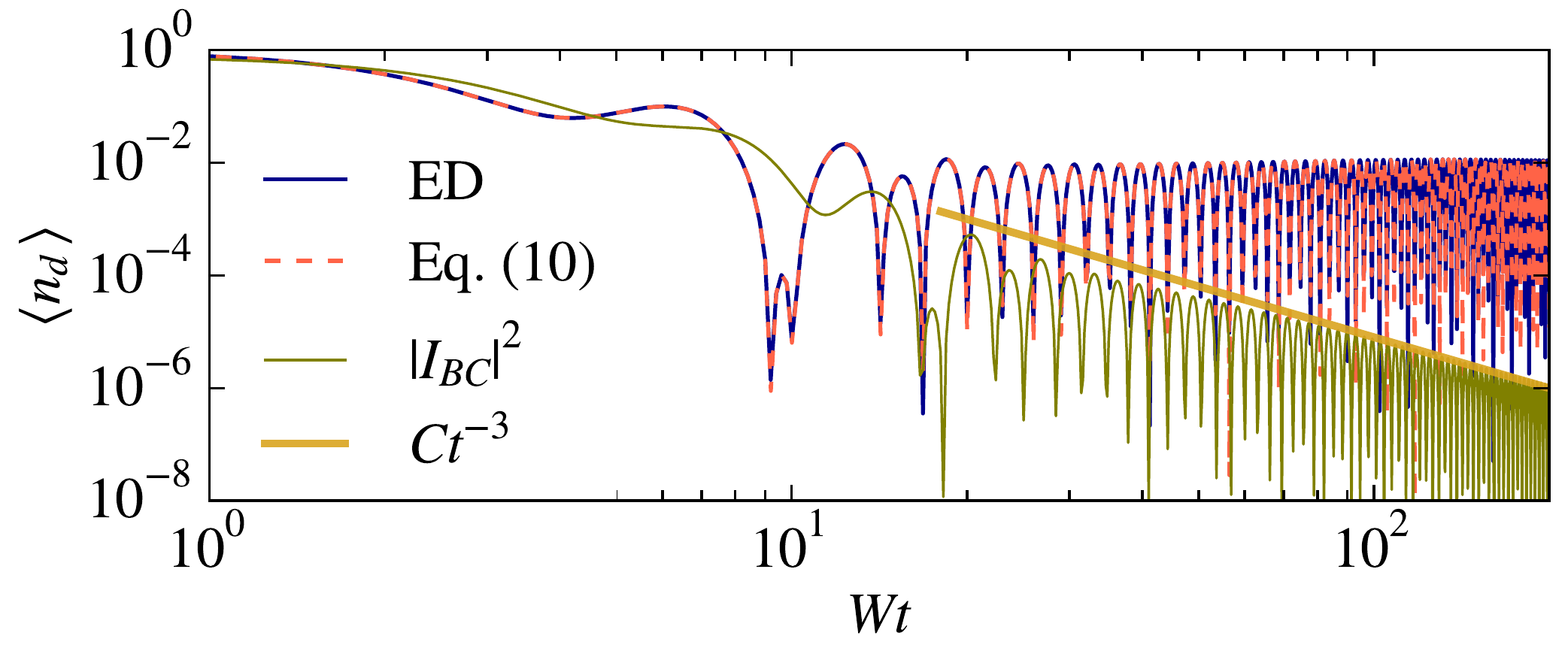}
\caption{Occupation probability $\braket{n_d(t)} = |\langle d | e^{-i H t} |d\rangle|^2$ as a function of time for an impurity coupled to a clean 1D lattice. The dark solid line is calculated by exact diagonalization (ED) for a chain of length $L = 1000$ and the red dashed line shows $\braket{n_d(t)}$ calculated by Eq.~(\ref{eq:wavefunction_amplitude}). The nonzero occupation at $t \rightarrow \infty$ is due to the two bound states [second term in Eq.~(\ref{eq:wavefunction_amplitude})]. The contribution of the continuum decays as $C t^{-3}$, where $C$ is a constant, as indicated. The model parameters are $E_d = 0$ and $g/W=0.5$.}
\label{fig:occupation_1D}
\end{figure}

We begin by checking numerically and illustrating the solution given in Eq.~(\ref{eq:wavefunction_amplitude}). We use exact diagonalization in 1D, whereas in higher dimension we use the Chebyshev expansion, Eq.~(\ref{eq:chebyshev_expansion}). Figure~\ref{fig:occupation_1D} shows $\braket{n_d(t)} = |\langle d|e^{-iHt}|d\rangle|^2$ calculated in the case of a clean 1D lattice. The result obtained with Eq.~(\ref{eq:wavefunction_amplitude}) using the known analytical expression of the impurity spectral function agrees with the solution by exact diagonalization and the long-time average equals the return probability given by Eq.~(\ref{eq:Qd1D}). The contribution of the continuum [first term in Eq.~(\ref{eq:wavefunction_amplitude})]
\begin{equation}
|I_{BC}|^2 = \left|\int_{E_{\min}}^{E_{\max}} dE\, e^{-i E t} A(E) \right|^2
\end{equation}
is shown separately. It decays as $t^{-3}$ because the continuum vanishes as a square root at the edges (see Fig.~\ref{fig:spectral_function} and Ref.~\onlinecite{Note1}). The long-time behavior of $\braket{n_d(t)}$ is therefore given by the second term due to the bound states.

\begin{figure}[tb]
\includegraphics[width=\linewidth]{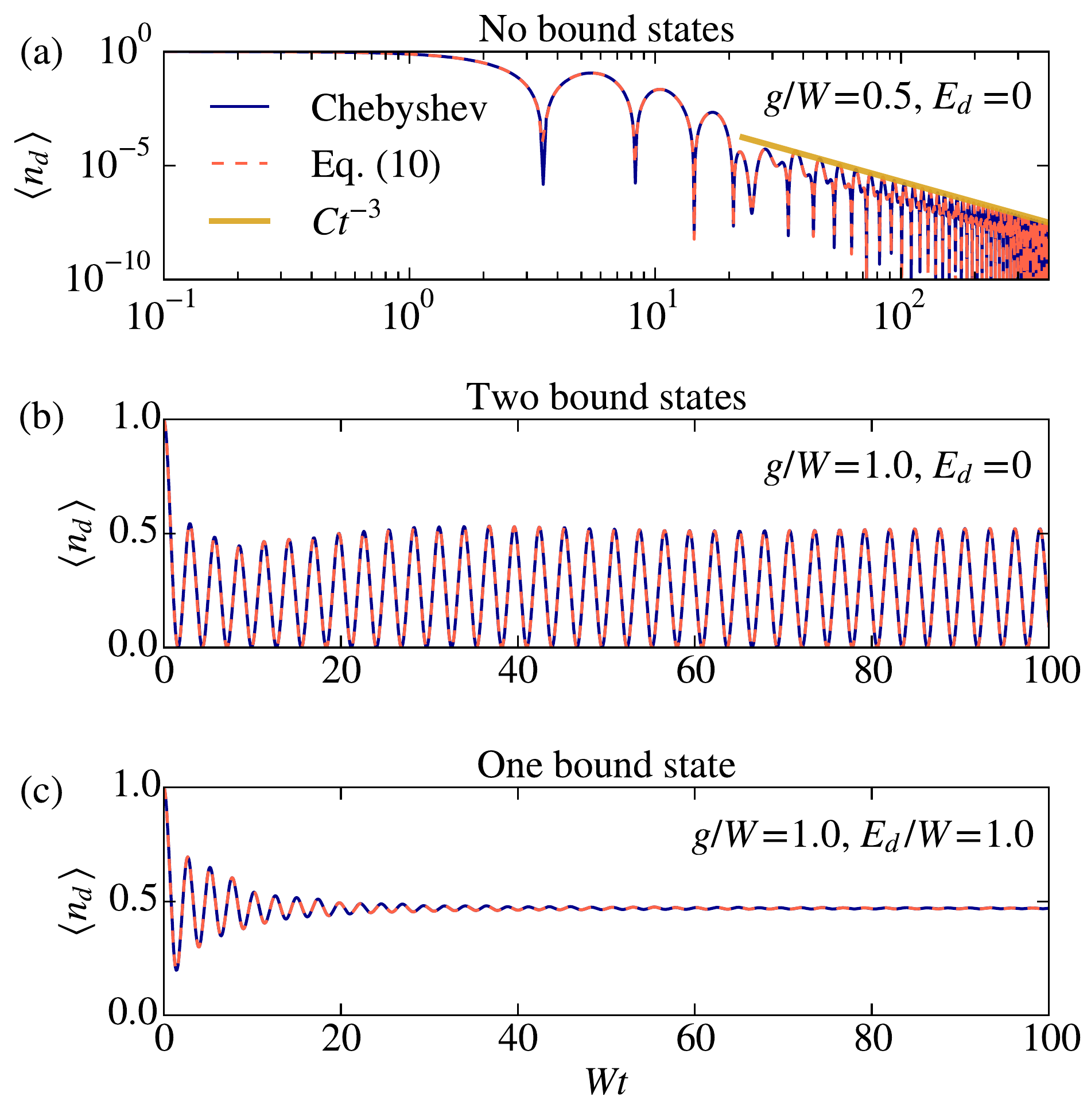}
\caption{Occupation probability as a function of time for an impurity coupled to a clean 3D lattice. Panels (a), (b), and (c) correspond to the solid, dashed, and dotted lines in Fig.~\ref{fig:intersections}, respectively. (a) When there are no bound states, $\langle n_d(t) \rangle$ is given by the first term in Eq.~(\ref{eq:wavefunction_amplitude}) and vanishes as $t^{-3}$. (b) Two bound states lead to Rabi oscillations as for a 1D lattice. (c) Only one bound state leads to saturation to a nonzero constant value. The Chebyshev expansion was truncated at order $M=1000$.
}
\label{fig:occupation_3D}
\end{figure}

The case of a 2D lattice is similar to the 1D case, with two bound states leading to a nonzero return probability. Additional structures develop as a function of time due to the Van Hove singularity in the lattice DOS, without consequences for the long-time behavior. For $D>2$, three qualitatively different evolutions can occur depending on the number of bound states. This is illustrated in Fig.~\ref{fig:occupation_3D} for $D=3$. For $g<g_c$, $\braket{n_d(t)}$ decreases as $t^{-3}$ due to the absence of bound state. Note that for $D\geqslant 3$, $A(E)$ vanishes at its edges like the lattice DOS with an exponent $D/2-1$, because $\Sigma_1(E)$ is finite at the edge. The two-bound states case shows Rabi oscillations like in 1D and 2D. Finally, in the case where only one bound state exists, $\braket{n_d(t)}$ approaches a constant value at long time.

\begin{figure}[tb]
\includegraphics[width=0.9\linewidth]{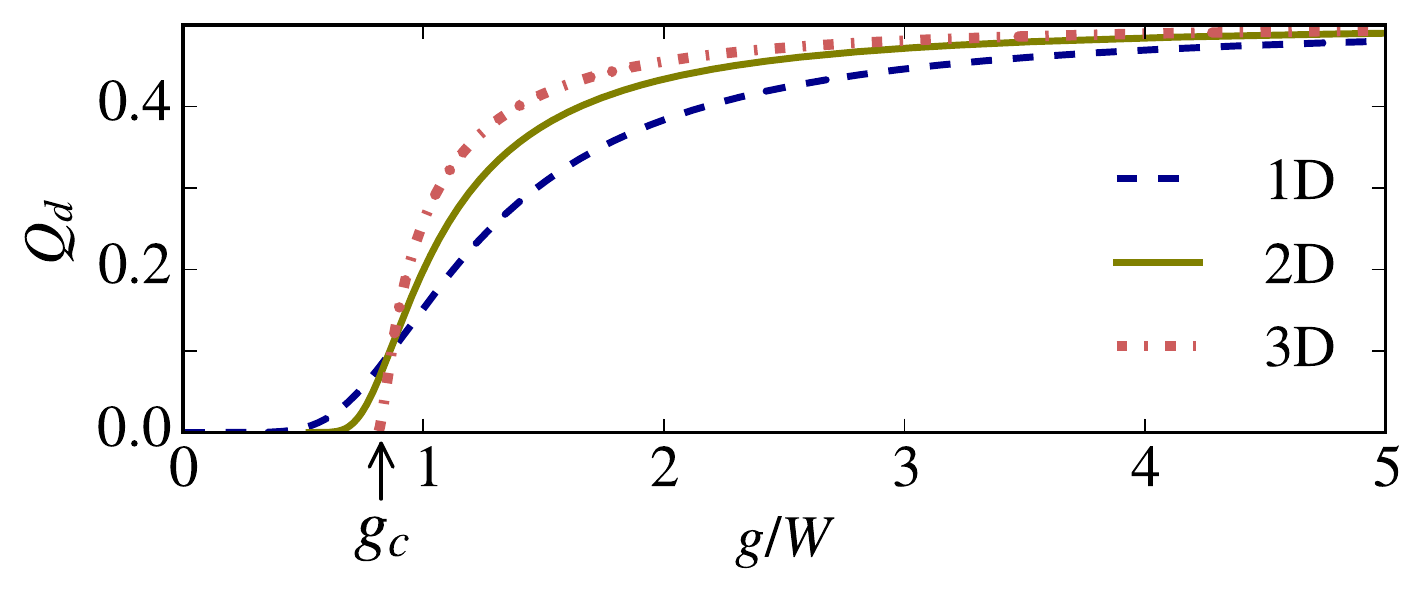}
\caption{Return probability as a function of $g$ for an impurity of energy $E_d=0$ coupled to a clean lattice. The 1D result is Eq.~(\ref{eq:Qd1D}) and the 2D and 3D curves are obtained numerically from Eqs.~(\ref{eq:return_probability_exact}) and (\ref{eq:bound_state_criterion}). In 3D, $Q_d$ vanishes at $g_c\approx 0.8\,W$ as indicated with an arrow. The return probability changes when the lattice is disordered, as shown in Figs.~\ref{fig:both_estimates} and \ref{fig:Qd_g_2D}.}
\label{fig:analytic_Qd}
\end{figure}

The evolution of the return probability with increasing impurity-lattice coupling is displayed in Fig.~\ref{fig:analytic_Qd} for the 1D, 2D, and 3D lattices. The return probability is discontinuous at $g = 0$: In 1D and 2D, $Q_d$ vanishes when $g$ approaches zero from above and in 3D it is identically zero, whereas $Q_d(g=0)=1$ in all dimensions. For a 3D lattice, there is a critical coupling $g_c$ at which bound states appear, such that $Q_d = 0$ for $0 < g < g_c$. In Fig.~\ref{fig:intersections}, $1/g_c^2$ is the slope of a line crossing $\Sigma_1(E)/g^2$ at the band edge. At large $g$, $Q_d$ approaches $1/2$ with a correction that decreases with increasing dimension, consistently with Eq.~(\ref{eq:Qd_clean_high_g}).

\subsection{Return probability in a disordered lattice}
\label{sec:disorder}

As seen in the previous sections, the return probability depends on the coupling $g$ due to possible bound states and is small at small $g$ (Fig.~\ref{fig:analytic_Qd}). In a disordered lattice, the bound states are modified and the existence of other localized states can lead to a large return probability even for small $g$. We investigate the effect of the coupling and disorder strength on the disorder-averaged return probability and the relation between $Q_d$ and the localization length in the lattice.

\subsubsection{One-dimensional lattice}

\begin{figure}[tb]
\includegraphics[width=\linewidth]{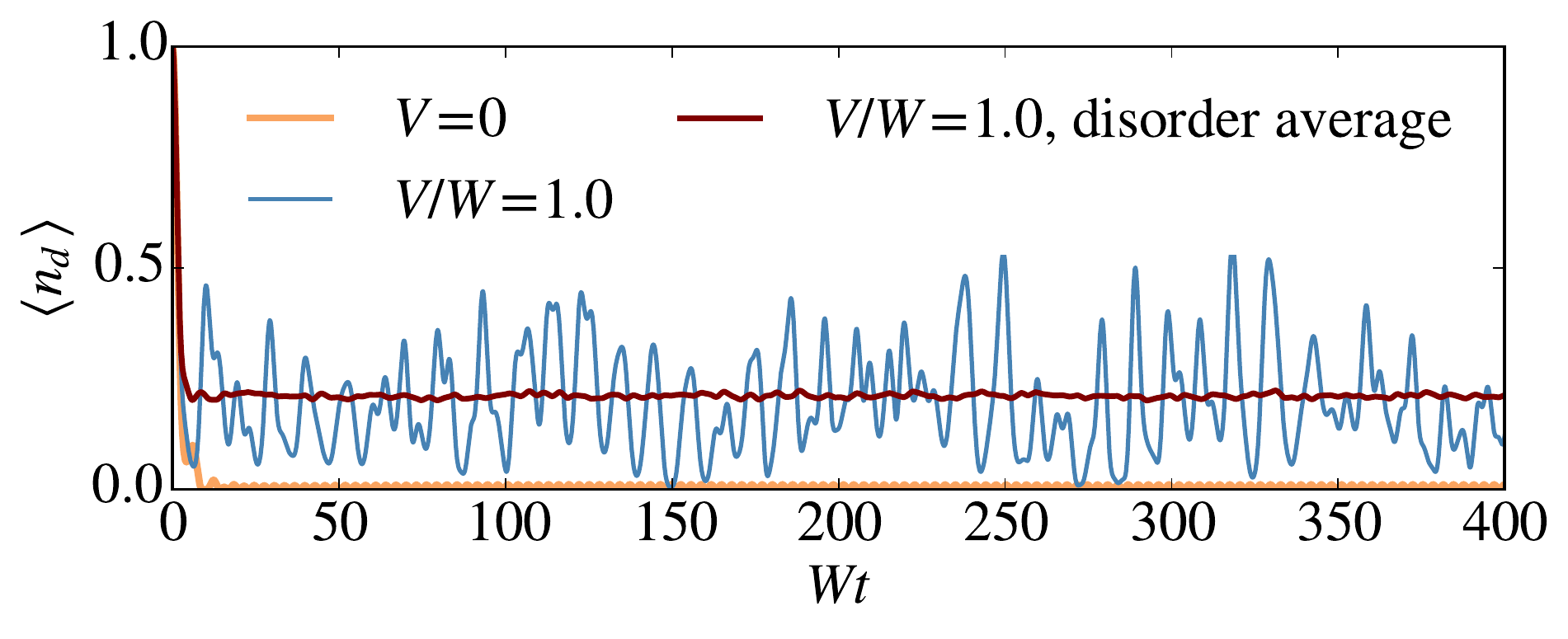}
\caption{Impurity occupation probability as a function of time for a clean (orange curve) and disordered (blue curve) one-dimensional lattice. The disorder increases the occupation probability, which oscillates with an irregular pattern. Upon averaging over 1000 realizations of the disorder, a well-defined return probability may be defined (red curve). The model parameters are $E_d = 0$ and $g/W=0.5$ with a chain of $L = 1000$ sites ruling out any finite-size effects within the simulation time.}
\label{fig:occupation_1D_disorder}
\end{figure}

\begin{figure}[b]
\includegraphics[width=\linewidth]{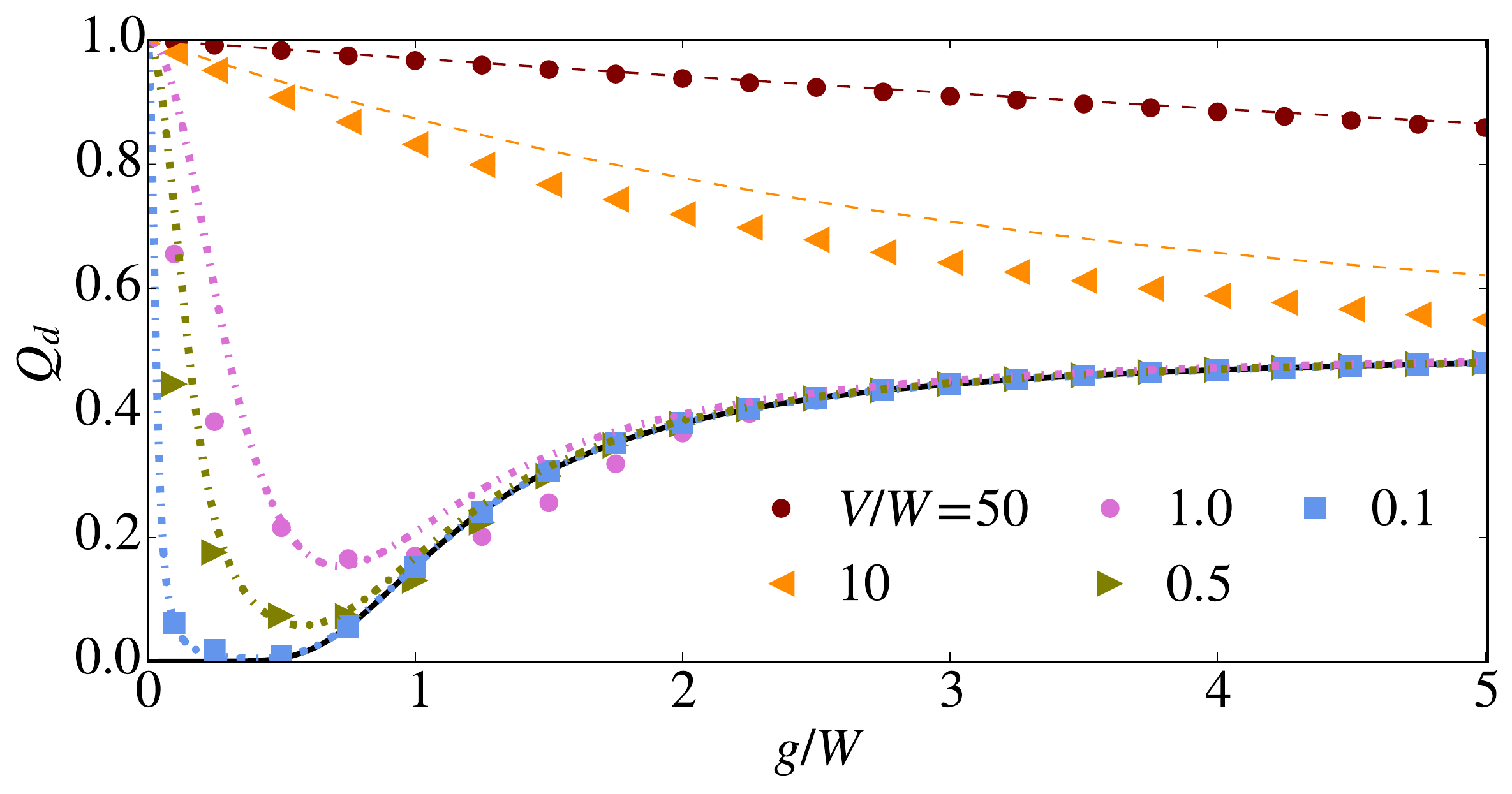}
\caption{Disordered-averaged return probability as a function of $g$ for an impurity of energy $E_d=0$ coupled to a 1D lattice. The markers show the numerical results for different disorder strengths $V$. The black solid line is the exact result for $V = 0$, Eq.~(\ref{eq:Qd1D}), the dashed lines show the strong-disorder model, Eq.~(\ref{eq:strong_disorder}), and the dash-dotted lines show Eq.~(\ref{eq:FGR_estimate}). The values of $\xi$ used in Eq.~(\ref{eq:FGR_estimate}) are 400, 17, and 4.5, corresponding to the disorder-averaged values calculated for $V = 0.1$, 0.5, and 1, respectively (see Appendix \ref{app:localization_length}). The lattice sizes used here are between $L = 1000$ and 5000, for which finite-size effects are negligible.}
\label{fig:both_estimates}
\end{figure}

The disorder-induced localization generally increases the return probability. A particle initially in the impurity state has an overlap with a certain number of localized states. The time evolution at long times is given by the oscillation between these localized states, which leads to an irregular oscillation of the occupation probability $\langle n_d(t) \rangle$, as seen in Fig.~\ref{fig:occupation_1D_disorder}. The occupation of an impurity state for a single realization of the disorder was studied in Ref.~\onlinecite{Garmon2017}. We consider here the return probability averaged over a large number $N$ of disorder realizations. For $N$ between 1000 and 2000, the results are well converged. The return probability is calculated by exact diagonalization using Eq.~(\ref{eq:return_probability}). Figure~\ref{fig:both_estimates} shows the disorder-averaged return probability as a function of the coupling. When there is no coupling, $Q_d = 1$, and the limiting value for $g/W \rightarrow \infty$ is $1/2$, as shown in Sec.~\ref{sec:strong_g}. For very strong disorder, $V=50\,W$, the points calculated by ED agree well with the model Eq.~(\ref{eq:strong_disorder}), which assumes that the impurity state is coupled to only one localized state of the lattice. The model overestimates $Q_d$ for $V=10\,W$. If $V>W$, the value of $Q_d$ is mostly set by the potential at the site $\vec{r}=\vec{0}$ (see Sec.~\ref{sec:strong_g}); when $|V_{\vec{0}}|$ is on average large, the impurity gets effectively decoupled and $Q_d$ approaches unity. In the intermediate to weak disorder regime, $V\lesssim W$, the value of $Q_d$ is controlled by the hybridization with the lattice: it first decreases from unity as $g$ increases, displays a minimum close to $g=V$, and approaches the value $1/2$ from below like in the clean case.

At weak coupling $g/W \lesssim 0.5$, the contribution of the bound states is negligible (Fig.~\ref{fig:analytic_Qd}). The finite value of $Q_d$ in this regime must therefore reflect the disorder in the lattice. The behavior of $Q_d$ may be explained qualitatively by considering a simplified model where the impurity state is coupled to the center of a box of length~$2\xi$. The impurity occupation is expected to decay exponentially over the time $T = 2 \xi/v$---during which the particle reaches the edge of the box and returns back to the origin---and this process is expected to repeat. An estimate of $Q_d$ is therefore given by the time average of the exponential decay over the time $T$. Here, $v$ is the group velocity at the impurity energy $E_d$. This simple model does not take into account the oscillations due to interferences between different eigenstates.
Since the decay rate is weakly affected by the disorder, we can use the Fermi golden rule value for the clean system, namely $\Gamma=2\pi g^2N(E_d)$. In 1D, the density of states $N(E_d)$ is simply $1/(\pi v)$. The return probability would then be
\begin{equation*}
\frac{1}{T} \int_0^T dt\, e^{-\Gamma t} = \frac{v^2}{4g^2\xi}\left(1 - e^{-4g^2\xi/v^2}\right).
\end{equation*}
This expression captures the behavior at small $g$, but not at large $g$ where the bound states dominate. An interpolation is obtained by adding their contribution given in Eq.~(\ref{eq:Qd1D}):
\begin{equation}
Q_d\approx\frac{v^2\left(1 - e^{-4g^2\xi/v^2}\right)}{4g^2\xi}+\frac{\left[ \sqrt{1 + 4(g/W)^4} - 1 \right]^2}{2 \left[1 + 4 (g/W)^4 \right]}.
\label{eq:FGR_estimate}
\end{equation}
In this simple model, the box size $2\xi$ corresponds to twice the localization length at energy $E=E_d=0$, which we calculate as the length $\xi_{\alpha}$ defined in Eq.~(\ref{eq:xi_alpha}) and averaged over the disorder at $E=0$, as described in Appendix~\ref{app:localization_length}. The result for $\xi = 400$, 17, and 4.5, corresponding to $V/W = 0.1$, 0.5, and 1, is plotted in Fig.~\ref{fig:both_estimates} and agrees reasonably well with the numerical solution.

Note that Fig.~\ref{fig:both_estimates} reveals a range of coupling $1 \lesssim g/W \lesssim 2$ where the inclusion of disorder slightly decreases the return probability with respect to the clean case. This counterintuitive result may be explained by the change of $\Sigma_1(E)$ with disorder. With increasing $V$, the divergences at the band edges become finite peaks which move outwards from the band edges. As seen in Eq.~(\ref{eq:return_probability_exact}), the return probability depends on the derivative $\partial_E \Sigma_1(E)|_{E = E_b}$. When the peaks shift outwards, the magnitude of the derivative at $E_b$ increases, leading to a smaller return probability. The energies $E_b$ also depend on the disorder but we expect that the average values of $E_b$ are unchanged.

We come now to our central question: Can the return probability $Q_d$, which is a local quantity, serve as a measure of the localization length $\xi$ in the disordered lattice? The localization length is a function of both energy and disorder strength. As shown in Appendix~\ref{app:localization_length}, different definitions of the localization length give slightly different but qualitatively consistent results: $\xi(E,V)$ is largest at $E=0$, which corresponds to the band center, and diverges as $1/V^2$ at small $V$. As $E$ approaches the band edges, $\xi$ drops to a value of the order of the lattice spacing. On the other hand, $Q_d(g,E_d,V)$ is a function of the coupling, the impurity energy, and the disorder strength. At weak coupling, the spectral function $A(E)$ shown in Fig.~\ref{fig:spectral_function} approaches a delta function at energy~$E_d$. One may expect that the value of $Q_d$ would then be set by the localized states close to $E=E_d$ and that $Q_d$ would be a function of $\xi(E_d,V)$ rather than $E_d$ and $V$ separately. In a hypothetical situation where $Q_d$ only depends on $\xi$ instead of all the parameters $g$, $E_d$, and $V$, there should be a universal relation between $Q_d$ and $\xi$ found by a proper scaling of the parameters. We find that this is only approximately true, and only provided that the contribution of the bound states to the return probability is negligible. We focus here on $V \lesssim W$, for which $\xi(E,V)$ varies from infinity to about two lattice spacings.

\begin{figure}[tb]
\includegraphics[width=\linewidth]{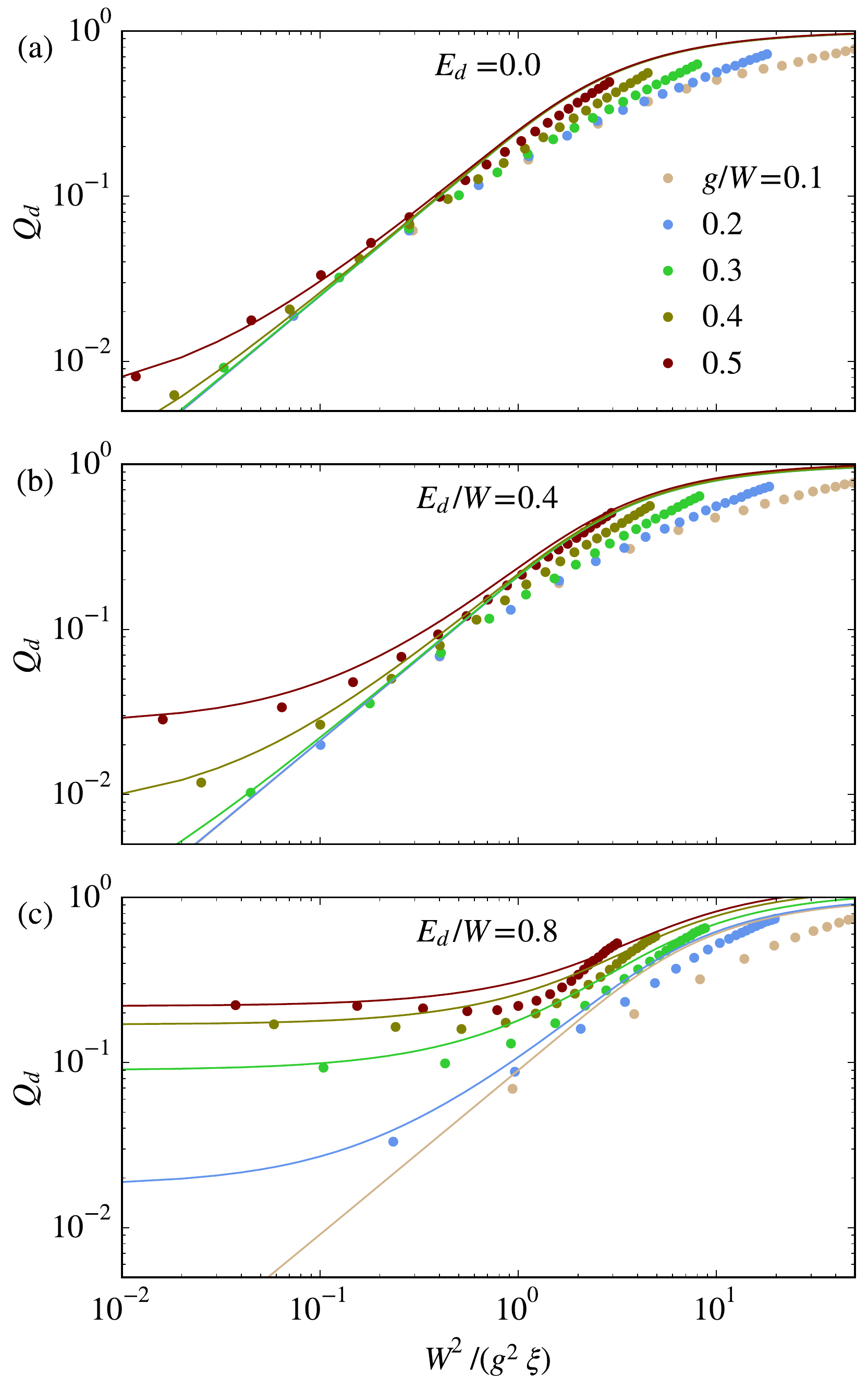}
\caption{Return probability $Q_d$ as a function of $W^2/(g^2\xi)$ calculated by varying $V$ and keeping $g$ and $E_d$ fixed. The dots are calculated by ED and the lines show (a) Eq.~(\ref{eq:FGR_estimate}) or (b, c) Eq.~(\ref{eq:FGR_estimate}) with the second term replaced by the corresponding numerical solution of Eqs.~(\ref{eq:bound_state_criterion}) and (\ref{eq:return_probability_exact}) for $E_d > 0$. For small $W^2/(g^2\xi)$ and $E_d$ close to the band center (a, b), the points for different values of $g$ approximately fall on the same curve, which is in good qualitative agreement with the model. When $E_d$ is close to the band edge (c), the larger contribution of bound states leads to a shift in $Q_d$ and the points deviate even at small $W^2/(g^2\xi)$. The leftmost markers correspond to $V/W = 0.1$ and the disorder increases from left to right in steps of $0.1$. The lattice size $L \gg \xi$ so that finite-size effects are negligible.}
\label{fig:scaling_1D}
\end{figure}

At small $g$ and $V \lesssim W$, Fig.~\ref{fig:both_estimates} shows an approximate agreement between the ED results and the first term of Eq.~(\ref{eq:FGR_estimate}). As the velocity $v/W = \sqrt{1 - (E_d/W)^2}$ is a constant for fixed $E_d$, this suggests that the return probability may become a universal function of $g^2 \xi$ in this regime. We test this hypothesis by plotting $Q_d$ as a function of $1/(g^2\xi)$ in Fig.~\ref{fig:scaling_1D}. We use for $1/\xi$ the value at $E=E_d$. For each value of the coupling $g$, we vary $1/\xi$ by sweeping the disorder strength between $0.1W$ and $2W$. With the impurity in the middle of the band, $E_d = 0$, Fig.~\ref{fig:scaling_1D}(a) shows that the points with varying couplings and disorder widths fall approximately on the same line when $1/(g^2\xi) \lesssim 0.5$, corresponding to a weak disorder and not too small a coupling. The scaling function seems to be slightly different from Eq.~(\ref{eq:FGR_estimate}), which is drawn as lines corresponding to each value of $g$; the agreement between the ED data and Eq.~(\ref{eq:FGR_estimate}) is best for the leftmost points corresponding to the weakest disorder or largest coupling.

The second term of Eq.~(\ref{eq:FGR_estimate}) results in a constant vertical shift between the different lines, which on a log-log scale is seen as a change of slope at small $1/(g^2 \xi)$. For $E_d = 0$, this shift is very small and the different lines mostly overlap.
For $E_d \neq 0$, the second term of Eq.~(\ref{eq:FGR_estimate}), which gives the contribution of the bound states, is modified as can be calculated numerically from Eqs.~(\ref{eq:bound_state_criterion}) and (\ref{eq:return_probability_exact}). A nonzero $E_d$ results in a larger weight of the bound states, which is seen as a larger vertical shift in Figs.~\ref{fig:scaling_1D}(b) and \ref{fig:scaling_1D}(c). Therefore, $Q_d$ depends separately on $E_d$ and is not only a function of $1/(g^2\xi)$ even for small values of $1/(g^2\xi)$. For $g/W = 0.5$ and $g/W = 0.4$, one can see that the combined effect of the bound states and disorder leads to a minimum in $Q_d$ instead of a monotonic increase: First, the weight of the bound states decreases due to disorder, creating a minimum of the return probability, and for large disorder the return probability increases again due to localization. Therefore, in order to measure $\xi$ via the return probability $Q_d$ in a regime where $Q_d$ does not depend separately on $V$, $g$, and $E_d$, one should choose $V \lesssim W$, $g/W \lesssim 0.5$, and $E_d$ close to the center of the band. In this regime, $Q_d$ is approximately proportional to $1/(g^2\xi)$, and one can effectively use the impurity as a probe of the localization length of the disordered lattice.

\subsubsection{Two-dimensional lattice}

\begin{figure}[b]
\includegraphics[width=\linewidth]{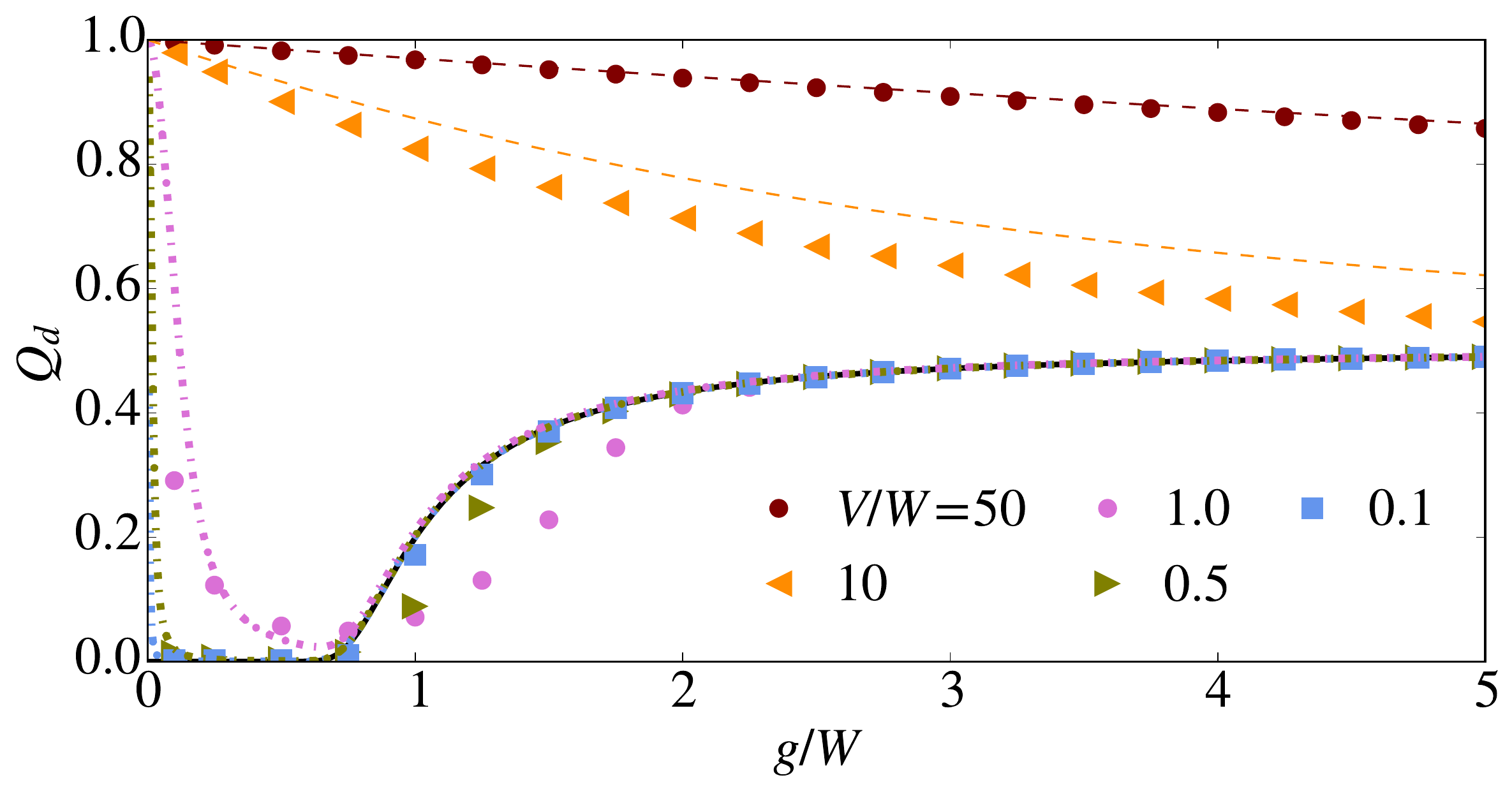}
\caption{The disorder-averaged return probability $Q_d$ as a function of $g$ as in Fig.~\ref{fig:both_estimates}, in the case of a 2D lattice. The black solid line for $V = 0$ is obtained from Eqs.~(\ref{eq:return_probability_exact}) and (\ref{eq:bound_state_criterion}), and the dashed lines show the strong-disorder result of Eq.~(\ref{eq:strong_disorder}). The dash-dotted lines show the estimate of Eq.~(\ref{eq:FGR_estimate_2D}) to which the $V = 0$ result has been added. The values of $\xi$ used in Eq.~(\ref{eq:FGR_estimate_2D}) are 5600, 350, and 8, corresponding to $V/W = 0.1$, 0.5, and 1.0, respectively, and are obtained by finite-size scaling (see Appendix~\ref{app:localization_length}). The lattice sizes used here are between $40 \times 40$ and $100 \times 100$. For $V/W = 0.1$ and 0.5, one can expect finite-size effects since $L < \xi$. They would however not be visible on the scale of the figure.}
\label{fig:Qd_g_2D}
\end{figure}

Differences can be expected when the impurity state is coupled to a 2D lattice, since the localization length is known to be much larger than in 1D. Figure~\ref{fig:Qd_g_2D} shows that for $V/W \lesssim 1$, $Q_d$ decreases faster as a function of $g/W$ than in 1D, indicating a smaller effect of localization. The estimates shown as dash-dotted lines are calculated in a similar way as in Fig.~\ref{fig:both_estimates}, albeit the values of $\xi$ used in the estimates are obtained by a finite-size scaling procedure, as explained in Appendix~\ref{app:localization_length}. We approximate the density of states by a constant, $N(E_d) \approx 1/(2 W)$, leading to the decay rate $\Gamma = \pi g^2/W$. For the velocity $v$, we use the average velocity of the constant-energy contour, which for $E = 0$ is $v = \sqrt{2} W/\pi$ and for other values of $E$ is calculated numerically. The first term of Eq.~(\ref{eq:FGR_estimate}) thus becomes
\begin{equation}
\frac{1}{T} \int_0^T dt\, e^{-\Gamma t} = \frac{W v}{2 \pi g^2 \xi} \left( 1 - e^{-\frac{2 \pi g^2 \xi}{W v}} \right).
\label{eq:FGR_estimate_2D}
\end{equation}
The contribution of the bound states, corresponding to the second term of Eq.~(\ref{eq:FGR_estimate}), is calculated numerically from Eqs.~(\ref{eq:bound_state_criterion}) and (\ref{eq:return_probability_exact}). This simple model agrees reasonably well with the numerical data. For $1 \lesssim g/W \lesssim 2$, the numerical data points fall more below the $V = 0$ analytic result than in 1D. This may be explained by a change in $\partial_E \Sigma_1(E)|_{E = E_b}$ as in 1D, which is more pronounced in 2D because $E_b$ is closer to the band edge for the same values of $g/W$. A disappearance of the bound states due to rounding of the band-edge singularity by disorder may also play a role. A definitive assessment would require us to identify in the numerics the bound states among the other discrete states of the disordered lattice and follow them as a function of $g$ and $V$, which is not straightforward.

\begin{figure}
\includegraphics[width=\linewidth]{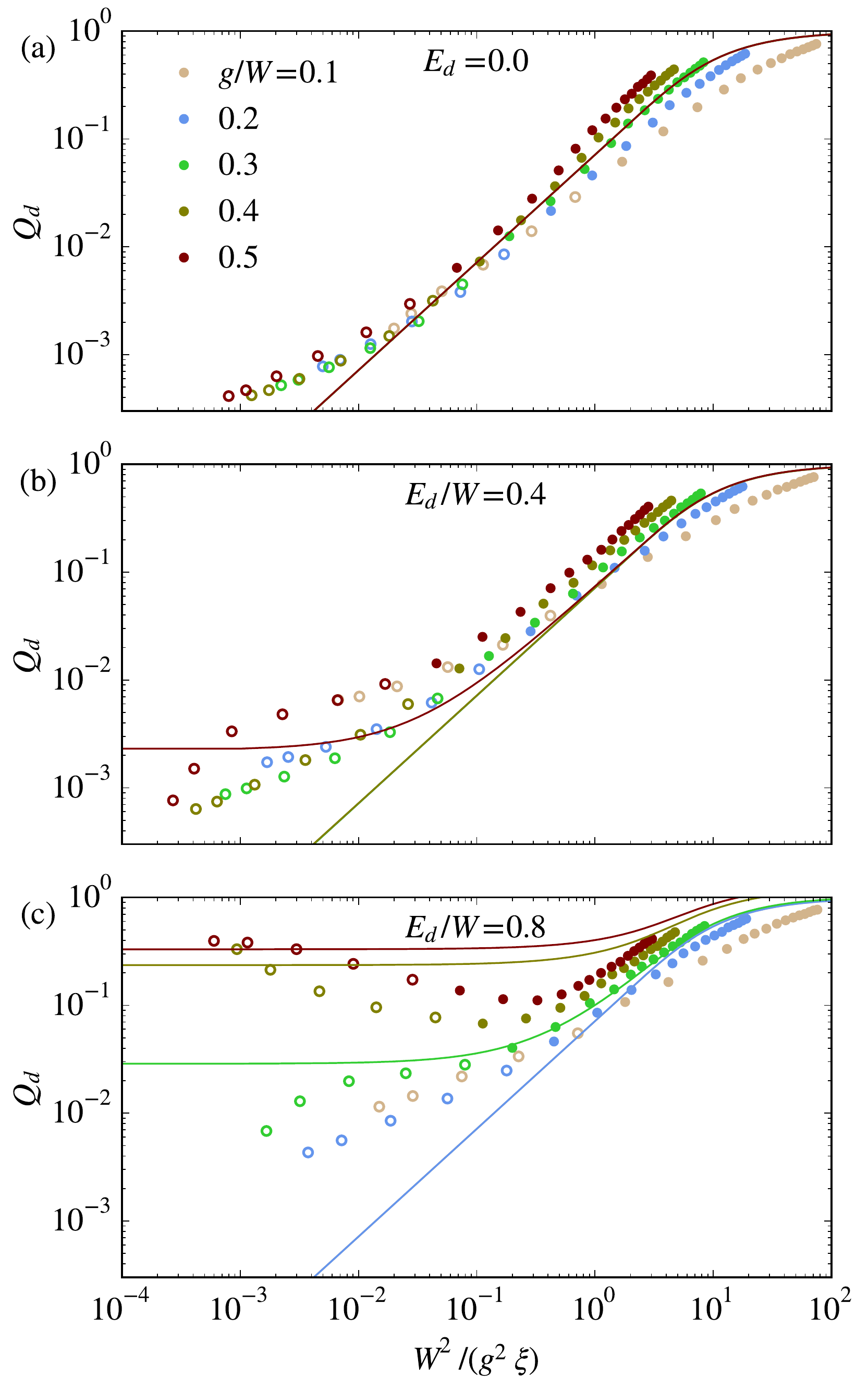}
\caption{Return probability $Q_d$ in the case of a 2D lattice, shown as a function of $W^2/(g^2\xi)$ (with $W = 4J$) as in Fig.~\ref{fig:scaling_1D}. The values of $V$ and $E_d$ are the same as in Fig.~\ref{fig:scaling_1D}. We use here values of $\xi$ obtained by finite-size scaling (Appendix~\ref{app:localization_length}). The hollow circles correspond to $\xi > L$, for which $Q_d$ is expected to depend on $L$ and therefore is size-specific. Here, $Q_d$ is calculated for a lattice of size $100 \times 100$ sites. At the center of the band ($E_d = 0$), the points for different couplings align approximately on the same line, whereas for larger $E_d/W$, there are more deviations between the points due to a larger contribution of the bound states. The solid lines given by the simple model of Eq.~(\ref{eq:FGR_estimate_2D}) and the bound-state contribution do not take into account the finite size of the lattice, which leads to deviations from the numerical results.}
\label{fig:scaling_2D}
\end{figure}

To analyze the dependence of $Q_d$ on $\xi$ in the case of a 2D lattice, we perform a finite-size scaling of $\xi$ as explained in Appendix~\ref{app:localization_length}. The return probability on the other hand is calculated for a specific size $L \times L$, and in the case of weak disorder depends on $L$. In Fig.~\ref{fig:scaling_2D}, points for which $\xi > L$ are marked with hollow circles to indicate that the results are size-specific, for a lattice of $100 \times 100$ sites. The simple model of Eq.~(\ref{eq:FGR_estimate_2D}) together with the bound-state contribution does not take into account the finite size of the lattice, which leads to differences between the numerical results and the model. For $E_d = 0$, the points for different $g$ fall approximately on the same line when $1/(g^2 \xi)$ is sufficiently small and $V\lesssim W$. Close to the band edge ($E_d/W = 0.8$), the bound states lead to large deviations between the points for different $g$ in the region of small $1/(g^2 \xi)$ where localization is weakest. The non-monotonic behavior of the return probability shows again the combined effect of the bound states and disorder. For $E_d/W=0.4$ and 0.8, the leftmost points with smallest $V$ show a slightly different trend than other points: this is a regime of disorder where $\xi$ exceeds $10^3$ lattice spacings and its precise value is uncertain (see Appendix~\ref{app:localization_length}).

\section{Discussion and possible experimental realizations}
\label{sec:experiments}

The previous sections highlight interesting phenomena occurring when an impurity level is coupled to a lattice in a regime where the Fermi golden rule is not applicable. For a clean lattice, the coupling can give rise to bound states outside the continuum and result in a nonzero return probability---i.e.\ a nonzero occupation of the impurity level---at infinite time. In dimensions one and two, two bound states necessarily arise below and above the lattice energy band such that persistent Rabi-like oscillations of the return probability survive at long times. The amplitude of these oscillations scales with a relatively high power of the impurity-lattice coupling and may be hard to detect at weak coupling. For a three- or higher-dimensional lattice, there can be two, one, or no bound states depending on the coupling and the energy of the impurity level relative to the band center. Each situation leads to a different behavior of the return probability at long time, namely oscillations, a saturation to a constant value, or a decay to zero, respectively.

For a disordered lattice, there are various regimes where the return probability is either dominated by the impurity-induced bound states like in the clean case, or by the Anderson localization of the lattice eigenstates. Strong disorder with respect to the lattice bandwidth eventually leads to eigenstates that are localized on a single lattice site, such that the impurity level is effectively coupled to only one state of the lattice. The disorder-averaged return probability in this limit can be understood by means of a two-level system involving the impurity and the localized state, provided that an average is made over the range of possible energies that the localized state can take. At strong impurity-lattice coupling, on the other hand, an effective two-level representation is again possible, which leads to an asymptotic return probability of $1/2$ with corrections scaling like the square of the disorder strength or bandwidth, whichever is largest.

In the most interesting case of weak coupling and low disorder, a simple model suggests that the return probability is a function of $g^2\xi$, where $g$ is the impurity-lattice coupling and $\xi$ is the localization length. We find that this is approximately true when the impurity level is close to the center of the band, both in 1D and 2D with similar behaviors. Hence in this particular regime---where (i) the coupling is weak enough and (ii) the impurity energy as far as possible from the bound states, such that the latter have negligible weight at the impurity, and (iii) the disorder is sufficiently low for the particle to have a chance of visiting the lattice---a measurement of the return probability yields information about the localization length in the lattice. When the impurity level is close to the band edges, however, the combined effect of the bound states and disorder leads to a non-monotonic and non-universal behavior of the return probability as a function of $g^2 \xi$.

Our results are for example relevant for experiments with ultracold atoms in optical potentials, where various quantum-mechanical models have been realized with a remarkable control over geometry and parameters.
In particular, various techniques exist for implementing disorder potentials. A recent experiment demonstrated a state-dependent laser speckle disorder potential\cite{Volchov_spectral_functions2018}. Radio-frequency coupling was used to transfer atoms from a Bose-Einstein condensate in a harmonic trap to another hyperfine state which feels the disorder \cite{Volchov_spectral_functions2018}. The model studied here could be realized with a similar scheme, using instead a local coupling and a state-dependent disorder potential in an optical lattice. The impurity state $\ket{d}$ would correspond to a hyperfine state that is unaffected by the disorder. Proposals have been made to use such local coupling for measuring the single-particle Green's function \cite{Kantian_lattice_assisted2015}. The single-atom and single-site precision required by these measurements is enabled by the recent development of quantum gas microscopy \cite{Kuhr_quantum-gas_microscopes2016}. Coupling an atom on a single site to a different hyperfine state \cite{Weitenberg_single-spin_addressing2011, Fukuhara_mobile_spin_impurity2013}, as well as disorder potentials \cite{Choi_many-body_localization2016}, have already been implemented in experiments with quantum gas microscopes \cite{Kuhr_quantum-gas_microscopes2016}. Furthermore, the digital micromirror device (DMD) allows one to create arbitrary potential landscapes for atoms \cite{Ott_single_atom_detection2016}. In an experiment which demonstrated the ``quantum walk'' of an atom, the DMD was used for creating an initial state of a single atom localized at one site of a one-dimensional lattice \cite{Preiss_quantum_walks2015}. As a realization of the model studied here, one could create a lattice with a side-attached impurity site as an alternative to locally coupling the atom to a different hyperfine state.

The combined presence of disorder and interactions can lead to strong modifications of the localization properties \cite{Altshuler_interaction_effects1980, Finkelstein_coulomb_interaction1983, Finkelstein_weak_localization1984, Altshuler_disordered_conductors1985, Lee_disordered1985, Giamarchi_localization_and_interaction1987, Giamarchi_Anderson_localization1988, Fisher_boson_localization1989, Belitz_Anderson-Mott1994}. Recently, the question of the ergodicity of such many-body localized states of interacting particles has been investigated in experiments \cite{Basko_metal_insulator2006, Kondov_strongly_correlated2015, Schreiber_observation_mbl2015, Bordia_coupling_identical2016, Choi_many-body_localization2016, Bordia_slow_relaxation2017, Luschen_slow_dynamics2017, Rubio-Abadal_quantum_bath2018, Lukin_entanglement2018}. How a finite density of interacting particles in the lattice affects the return probability to an impurity level remains an open problem.

\section{Conclusions}
\label{sec:conclusions}

In this paper, we have investigated the return probability of a particle to an impurity level hybridized with a clean or disordered lattice. We have shown that depending on dimension, bound states can emerge and lead to a nonzero return probability even in a clean lattice. For disordered lattices, different regimes of the hybridization and disorder strength lead to different behaviors of the return probability with nontrivial effects of the bound states and disorder combined. We have investigated the possibility of using the return probability, which is an out-of-equilibrium \emph{local} observable, as a probe of the localization length in the lattice, which is a non-local property. In short, the return probability can provide a useful measure of the localization length for 1D and 2D lattices in the regime $E_d\approx0$, $g\lesssim W/2$, and $V\lesssim W$, where $E_d$ is the impurity energy measured from the center of the lattice energy band, $2W$ is the bandwidth, $g$ is the impurity-lattice coupling, and $V$ is the strength of disorder. In this regime, the return probability is roughly proportional to $1/(g^2 \xi)$, where $\xi$ is the localization length at the energy $E_d$.

The present study deals with an impurity level coupled to the simplest bath, that is an empty lattice. A first step to extend the study to more complex baths would be to consider a bath occupied by a finite density of particles. In the clean case, this would allow one to study effects such as the Anderson orthogonality catastrophe \cite{Anderson_infrared_catastrophe1967, Anderson_ground_state1967}. In the case of a disordered interacting bath, an interesting question is whether a particle in an impurity state could be used as a probe of many-body localization.

\begin{acknowledgments}
This work was supported by the Wihuri foundation and in part by the Swiss National Science Foundation under Division II and by the ARO-MURI Non-equilibrium Many-body Dynamics grant (W911NF-14-1-0003). The calculations were performed in the University of Geneva with the clusters Mafalda and Baobab.
\end{acknowledgments}

\appendix

\section{Lattice return probability}
\label{app:return_probability}

Localization can be measured by the return probability to a given initial state $\ket{\psi}$. The probability of remaining in the state $\ket{\psi}$ after a time $t$ is
\begin{align*}
\langle n_{\psi}(t) \rangle &= |\bra{\psi} e^{-iH_0t} \ket{\psi}|^2 \\
&= \sum_{\alpha\beta}\bra{\psi} e^{-iH_0t} \ket{\alpha}\braket{\alpha|\psi} \bra{\psi} e^{iH_0t}\ket{\beta} \braket{\beta | \psi} \\
&= \sum_{\alpha\beta} e^{-i(E_{\alpha} - E_{\beta})t} |\braket{\psi|\alpha}|^2 |\braket{\psi|\beta}|^2,
\end{align*}
where $\ket{\alpha}$ and $\ket{\beta}$ are eigenstates of $H_0$. The return probability is the long-time limit of the time-averaged probability:
\begin{align*}
Q_{\psi} &= \lim_{T \rightarrow \infty} \frac{1}{T} \int_0^T \langle n_{\psi}(t) \rangle dt \\
&= \sum_{\alpha\beta}|\braket{\psi|\alpha}|^2 |\braket{\psi|\beta}|^2 \lim_{T \rightarrow \infty} \frac{1}{T}\int_0^T e^{-i(E_{\alpha} - E_{\beta})t}\\
&= \sum_{\alpha\beta}|\braket{\psi|\alpha}|^2 |\braket{\psi|\beta}|^2 \delta_{\alpha\beta}
= \sum_{\alpha} |\braket{\psi|\alpha}|^4.
\end{align*}
The last line holds under the assumption of non-degenerate energies $E_{\alpha}$. If the initial state is a position eigenstate $|\vec{r}\rangle$ and after performing a disorder average, we arrive at Eq.~(\ref{eq:Qr_eigenstates}).

\section{Solution by Laplace transform}
\label{app:analytic_Laplace}

We project the wave function $\ket{\psi(t)}=e^{-iHt}\ket{d}$ on the basis formed by the impurity state $\ket{d}$ and the lattice eigenstates $\ket{\alpha}$ with energies $E_{\alpha}$:
\begin{equation}\label{eq:projection}
\ket{\psi(t)} = e^{-i E_d t} \psi_d(t) \ket{d} + \sum_{\alpha} e^{-i E_{\alpha} t}\psi_{\alpha}(t) \ket{\alpha}.
\end{equation}
In this basis, the Hamiltonian given by Eq.~(\ref{eq:Hamiltonian}) is $H=\sum_{\alpha}E_{\alpha}|\alpha\rangle\langle\alpha|+E_d|d\rangle\langle d|+g\sum_{\alpha}\big(\langle\vec{0}|\alpha\rangle|d\rangle\langle\alpha|+\mathrm{H.c.}\big)$. The Schr{\"o}dinger equation $i\partial_t\ket{\psi(t)}=H \ket{\psi(t)}$ becomes
\begin{align}
\label{eq:psid}
i\partial_t \psi_d(t) &=g\sum_{\alpha}\braket{\vec{0} | \alpha} e^{i(E_d - E_{\alpha}) t} \psi_{\alpha}(t) \\
\label{eq:psialpha}
i\partial_t \psi_{\alpha}(t) &= g \braket{\alpha | \vec{0}} e^{-i(E_d - E_{\alpha}) t} \psi_d(t).
\end{align}
This is to be solved with initial condition $\psi_d(0) = 1$ and $\psi_{\alpha}(0) = 0$. Integration of Eq.~(\ref{eq:psialpha}) and substitution in Eq.~(\ref{eq:psid}) yields
\begin{align}
\nonumber
\partial_t \psi_d(t) &= -\int_0^t dt'\,\psi_d(t')g^2\sum_{\alpha}|\langle\vec{0}|\alpha\rangle|^2 e^{i(E_d - E_{\alpha})(t-t')} \\
&\equiv -\int_0^t dt' \psi_d(t') M(t-t'),
\label{eq:diff_eq}
\end{align}
where we have defined the memory function as
\begin{equation}
M(t) = g^2\sum_{\alpha}|\langle\vec{0}|\alpha\rangle|^2 e^{i(E_d-E_{\alpha})t}.
\label{eq:self-energy_d}
\end{equation}
The Laplace transformation is well-suited for initial-value problems like Eq.~(\ref{eq:diff_eq}). We recall the main properties of this transformation for clarity. The Laplace transform and inverse transform of a function $f(t)$ are defined as
\begin{align*}
\mathscr{L}[f(t)] = \tilde{f}(z) &= \int_0^{\infty} dt\, f(t) e^{-z t} \\
f(t) &= \frac{1}{2 \pi i} \int_{\delta - i \infty}^{\delta + i \infty} dz\, \tilde{f}(z) e^{z t},
\end{align*}
where $z \in \mathbb{C}$ and $\delta \in \mathbb{R}$ lies on the right side of all singularities of $\tilde{f}(z)$. The derivative and convolution have simple Laplace transforms similar to their Fourier transforms:
\begin{align*}
\mathscr{L}\left[\partial_t f(t)\right] &= z \tilde{f}(z) - f(0) \\
\mathscr{L}\left[\int_0^t dt' f(t') g(t - t')\right] &= \tilde{f}(z) \tilde{g}(z).
\end{align*}
The transformation of Eq.~(\ref{eq:diff_eq}) gives the algebraic equation $z \tilde{\psi}_d(z) - 1 = -\tilde{\psi}_d(z) \tilde{M}(z)$ with the solution
\begin{equation}\label{eq:solution1}
\tilde{\psi}_d(z) = \frac{1}{z + \tilde{M}(z)}.
\end{equation}
The transformation of Eq.~(\ref{eq:self-energy_d}) gives $\tilde{M}(z)$ as
\begin{equation}\label{eq:solution2}
\tilde{M}(z) = ig^2\sum_{\alpha}\frac{|\langle\vec{0}|\alpha\rangle|^2}{iz + E_d - E_{\alpha}}=i\Sigma(iz+E_d),
\end{equation}
where we took advantage of the analytic continuation of the impurity self-energy defined in Eq.~(\ref{eq:self-energy}) into the complex plane:
\begin{equation}
\Sigma(z) = g^2\sum_{\alpha} \frac{|\langle\vec{0}|\alpha\rangle|^2}{z -E_{\alpha}}.
\label{eq:self-energy_new}
\end{equation}
Taking the inverse transform of Eq.~(\ref{eq:solution1}) and using Eq.~(\ref{eq:solution2}), the amplitude on the impurity level can now be written as a line integral in the complex plane,
\begin{equation}\label{eq:solution3}
\psi_d(t) = \frac{1}{2 \pi i} \int_{\delta - i \infty}^{\delta + i \infty} dz\,\frac{e^{zt}}{z + i \Sigma(iz + E_d)}.
\end{equation}
All singularities of the integrand lie on the imaginary axis, such that one can set $\delta=0^+$. This can be seen by rewriting the equation $z+i\Sigma(iz+E_d)=0$ in the form
	\[
		z+g^2\sum_{\alpha}\left[z^*+i(E_d-E_{\alpha})\right]\frac{|\langle\vec{0}|\alpha\rangle|^2}{|z-i(E_d-E_{\alpha})|^2}=0,
	\]
which shows that all solutions have $\mathrm{Re}\,z=0$. It is convenient to change variable from $z$ to $z'=iz+E_d$ in Eq.~(\ref{eq:solution3}), which rotates the integration line to just above the real axis and gives:
\begin{equation*}
\psi_d(t) = -\frac{e^{iE_dt}}{2 \pi i} \int_{-\infty + i \delta}^{\infty + i \delta} dz'\,\frac{e^{-iz't}}{z' - E_d - \Sigma(z')}.
\end{equation*}
The phase factor cancels the one in Eq.~(\ref{eq:projection}), such that the time-dependent occupation amplitude of the impurity level is
\begin{equation}
\braket{d | e^{-i H t} | d} = -\frac{1}{2 \pi i} \int_{-\infty + i \delta}^{\infty + i \delta} dz\,\frac{e^{-izt}}{z - E_d - \Sigma(z)}.
\label{eq:complex_integral}
\end{equation}
The integrand can be identified as the Green's function of the impurity level, $G_{dd}(z)=1/[z-E_d-\Sigma(z)]$, whose spectral function $A(E)=(-1/\pi)\mathrm{Im}\,G_{dd}(z\to E+i0)$ is given by Eqs.~(\ref{eq:impurity_spectral_function}) and (\ref{eq:self-energy}). $G_{dd}(z)$ has singularities on the real axis, including the continuum extending from $E_{\min}$ to $E_{\max}$---this becomes a quasi-continuum for a finite or sufficiently disordered system---and the possible bound states outside the continuum.

\begin{figure}[tb]
\includegraphics[width=0.8\linewidth]{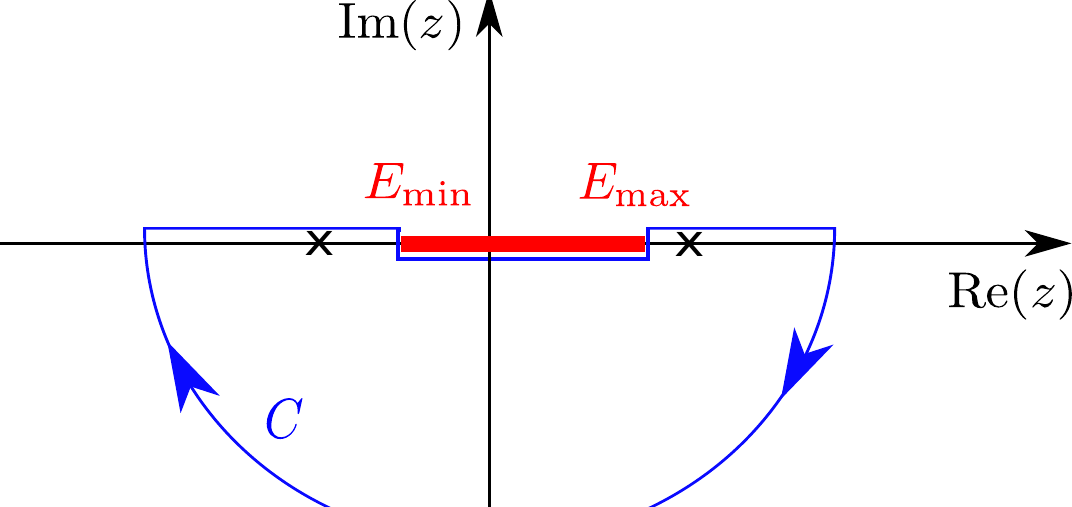}
\caption{The integration contour $C$ of Eq.~(\ref{eq:contour_part}). The branch cut between $E_{\min}$ and $E_{\max}$ is marked with red color and the possible poles on the real axis with crosses.}
\label{fig:integration_contour}
\end{figure}

Equation (\ref{eq:complex_integral}) is transformed into Eq.~(\ref{eq:wavefunction_amplitude}) by means of the residue theorem. Due to the factor $e^{-izt}$ and the fact that $t>0$, we must close the contour in the lower-half of the complex plane as illustrated in Fig.~\ref{fig:integration_contour}, avoiding the interval between $E_{\min}$ and $E_{\max}$ such that the integrand is analytic inside the contour, and correcting with the difference between the value of the integrand above and below that interval. The integral in Eq.~(\ref{eq:complex_integral}) thus becomes a sum of two parts, $\braket{d | e^{-i H t} | d} =I_C + I_{BC}$, where $I_C$ is the contribution of the contour, which yields the residues at the bound states:
\begin{multline}
I_C = -\frac{1}{2 \pi i} \oint_C dz \frac{e^{-i z t}}{z - E_d - \Sigma(z)} \\
= \sum_{E_b} \frac{e^{-i E_b t}}{1 - \partial_E \Sigma_1(E)|_{E = E_b}}.
\label{eq:contour_part}
\end{multline}
We use the notation $\Sigma(E)\equiv\Sigma(z\to E+i0)$ for the retarded self-energy evaluated just above the real axis and we have used the condition that $\Sigma_2(E)$ must vanish at the energy of the bound-states. Since the contour $C$ goes under the real axis for $\text{Re}\,z$ between $E_{\min}$ and $E_{\max}$, we must subtract this part and add the part above the real axis. The second term $I_{BC}$ is therefore
	\begin{align}\label{eq:branch_cut_part}
		\nonumber
		I_{BC}&=-\frac{1}{2\pi i}\int_{E_{\min}}^{E_{\max}}\hspace{-0.8em}dE\,e^{-iEt}
		\left[G_{dd}(E+i0)-G_{dd}(E-i0)\right]\\
		&=\int_{E_{\min}}^{E_{\max}}dE\,e^{-iEt}A(E).
	\end{align}
We have used the property $\Sigma(z^*)=\Sigma^*(z)$ [see Eq.~(\ref{eq:self-energy_new})], which implies $G_{dd}(z^*)=G_{dd}^*(z)$ and consequently $G_{dd}(E+i0)-G_{dd}(E-i0)=2i\mathrm{Im}\,G_{dd}(E+i0)=-2\pi iA(E)$. The sum of Eqs.~(\ref{eq:contour_part}) and (\ref{eq:branch_cut_part}) yields Eq.~(\ref{eq:wavefunction_amplitude}).

\section{Solution by equation of motion}
\label{app:analytic_Greens_function}

The solution by Laplace transform uses the ``Schr{\"o}\-din\-ger picture'' with time-dependent wave function, while the equation of motion method for the Green's function is based on the Heisenberg picture with time-dependent operators. The latter is more easily generalized to a many-particle context. We describe it here for fermions. The retarded Green's function of interest in our case is
\begin{equation}
G_{\mu \nu}(t) = -i \Theta(t) \braket{d | \left[ c_{\mu}^{\phantom{\dagger}}(t), c_{\nu}^{\dagger}(0) \right]_+ | d},
\label{eq:greens_function}
\end{equation}
where $\mu, \nu \in \{ d, \alpha \}$, $\Theta(t)$ is the Heaviside function, the operators evolve in time according to $c_{\mu}(t) = e^{i H t} c_{\mu} e^{-i H t}$, and $[\cdot\,,\cdot]_+$ is the anti-commutator. Because $c^{\dagger}_d$ destroys the state $\ket{d}$, one sees that
	\begin{equation}\label{eq:Green1}
		\braket{d | e^{-i H t} | d} =iG_{dd}(t)
	\end{equation}
for $t>0$. The equation of motion of $G_{\mu \nu}(t)$ is
\begin{equation}\label{eq:Green2}
	i\partial_t G_{\mu\nu}(t) =\delta_{\mu\nu}\delta(t)-i\Theta(t)
	\braket{d | \left[ \left[c_{\mu}(t),H\right], c_{\nu}^{\dagger} \right]_+ | d},
\end{equation}
where the first term on the right-hand-side comes from differentiating the Heaviside function and using the anti-commutation rule $\left[c_{\mu}^{\phantom{\dagger}},c_{\nu}^{\dagger}\right]_+=\delta_{\mu\nu}$ and the second term uses the equation of motion of the operators, $\partial_t c_{\mu}(t)=-i\left[c_{\mu}(t),H\right]$. Expressed in terms of the $c_d$ and $c_{\alpha}$, the Hamiltonian is $H=\sum_{\alpha}E^{\phantom{\dagger}}_{\alpha}c^{\dagger}_{\alpha}c^{\phantom{\dagger}}_{\alpha}+E^{\phantom{\dagger}}_dc^{\dagger}_dc^{\phantom{\dagger}}_d+g\sum_{\alpha}\left(\langle\vec{0}|\alpha\rangle c^{\dagger}_dc^{\phantom{\dagger}}_{\alpha}+\mathrm{H.c.}\right)$. We deduce the commutators entering Eq.~(\ref{eq:Green2}),
\begin{align*}
[c_d, H] &= E_d c_d + g\sum_{\alpha}\langle\vec{0}|\alpha\rangle c_{\alpha}\\
[c_{\alpha}, H] &= E_{\alpha} c_{\alpha} + g\langle\alpha|\vec{0}\rangle c_d,
\end{align*}
and obtain two coupled equations for $G_{dd}$ and $G_{\alpha d}$ that are the counterpart of Eqs.~(\ref{eq:psid}) and (\ref{eq:psialpha}):
\begin{align}
i\partial_t G_{dd}(t) &=\delta(t)+E_dG_{dd}(t)+g\sum_{\alpha}\langle\vec{0}|\alpha\rangle G_{\alpha d}(t)\\
i\partial_t G_{\alpha d}(t) &=E_{\alpha}G_{\alpha d}(t)+g\langle\alpha|\vec{0}\rangle G_{dd}(t).
\end{align}
Fourier transforming these equations from $t$ to $\omega$ and continuing analytically to the complex plane $\omega\to z$ yields
\begin{align}
(z-E_d)G_{dd}(z) &=1+g\sum_{\alpha}\langle\vec{0}|\alpha\rangle G_{\alpha d}(z)\\
(z-E_{\alpha})G_{\alpha d}(z) &=g\langle\alpha|\vec{0}\rangle G_{dd}(z),
\end{align}
with the solution
\begin{equation}
	G_{dd}(z)=\frac{1}{z-E_d-\Sigma(z)},
\end{equation}
where $\Sigma(z)$ is defined in Eq.~(\ref{eq:self-energy_new}). The function $G_{dd}(z)$ is analytic in the upper half of the complex plane and vanishes as $1/z$ for $z\to\infty$. These conditions are sufficient for the Fourier transform of $G(z\to E+i0)$ to be proportional to $\Theta(t)$ as required by Eq.~(\ref{eq:greens_function}). We therefore have
\begin{equation}
	G_{dd}(t)=\int_{-\infty+i0}^{\infty+i0}\frac{dz}{2\pi}\,\frac{e^{-izt}}{z-E_d-\Sigma(z)},
\end{equation}
which, on account of Eq.~(\ref{eq:Green1}), is just Eq.~(\ref{eq:complex_integral}).

\section{Chebyshev expansion}
\label{app:chebyshev}

The Chebyshev polynomials $T_m(x)=\cos(m\arccos x)$ with integer $m\geqslant 0$ form a basis for representing functions $f(x)$ having support in the interval $-1<x<1$. The expansion reads $f(x)=\sum_{m=0}^{\infty}c_mT_m(x)$ with coefficients given by
	\begin{equation}
		c_m=\frac{2-\delta_{m0}}{\pi}\int_{-1}^{1}dx\,\frac{f(x)T_m(x)}{\sqrt{1-x^2}}.
	\end{equation}
In order to find the expansion of the evolution operator, we consider the function $e^{-ixt}$ for $1<x<1$, change variable from $x$ to $\vartheta$ with $x=\cos\vartheta$, use the representation $e^{-it\cos\vartheta}=\sum_{n=-\infty}^{\infty}(-i)^nJ_n(t)e^{-in\vartheta}$, where $J_n$ are the Bessel functions of the first kind, as well as the property $J_{-n}(t)=(-1)^nJ_n(t)$, to arrive at $c_m=\left(2-\delta_{m0}\right)(-i)^mJ_m(t)$. It follows that
	\begin{equation}
		e^{-ixt}=\sum_{m=0}^{\infty}\left(2-\delta_{m0}\right)(-i)^mJ_m(t)T_m(x).
	\end{equation}
Replacing $x$ by $H=b+a\tilde{H}$ on the left-hand side, we deduce Eq.~(\ref{eq:chebyshev_expansion}). For expanding the Green's function, we consider $f(x)=1/(z-x)$ with $z\in\mathbb{C}$ and proceed with the same change of variable. We then use the identity
	\begin{equation*}
		\int_0^{\pi}d\vartheta\,\frac{\cos(m\vartheta)}{z-\cos\vartheta}
		=\frac{-i\pi e^{-im\arccos z}}{\sqrt{1-z^2}}\qquad(\mathrm{Im}\,z>0)
	\end{equation*}
to arrive at an expression valid for $z$ in the upper half of the complex plane:
	\begin{equation}
		\frac{1}{z-x}=\sum_{m=0}^{\infty}\frac{i\left(\delta_{m0}-2\right)e^{-im\arccos z}}{\sqrt{1-z^2}}
		T_m(x).
	\end{equation}
The expansion of $1/(E+i0-H_0)$ follows and takes the form given in Eq.~(\ref{eq:chebyshev_expansionG}).

A calculation of the time-dependent impurity-level occupation based on Eq.~(\ref{eq:chebyshev_expansion}) or a calculation of the lattice Green's function based on Eq.~(\ref{eq:chebyshev_expansionG}) reduces to the evaluation of the matrix elements $\langle d|T_m(\tilde{H})|d\rangle$ or $\langle\vec{r}|T_m(\tilde{H}_0)|\vec{r}'\rangle$. This is greatly simplified thanks to the recursion relation $T_m(x)=2xT_{m-1}(x)-T_{m-2}(x)$ obeyed by the Chebyshev polynomials: rather than evaluating high-order polynomials of the Hamiltonian, one uses an iterative procedure by applying the Hamiltonian repeatedly. As the storage of the Hamiltonian matrix in the computer memory is not required, this opens the way for treating systems of very large size.

\section{Return probability for strong disorder}
\label{app:strong_disorder}

The two-level subsystem formed by the impurity and the disordered-lattice eigenstate localized at $\vec{r}=\vec{0}$ is described by the $2\times2$ Hamiltonian
\begin{equation*}
\begin{pmatrix}
E_d	&g \\
g	&E_0
\end{pmatrix},
\end{equation*}
where $E_0$ is the energy of the localized state in the range $|E_0|<W+V$. The eigenvalues are
\begin{equation*}
E_{\pm} = \frac{E_d + E_0}{2} \pm \sqrt{\left(\frac{E_d - E_0}{2}\right)^2 + g^2}
\end{equation*}
and the eigenvectors can be written as
\begin{align*}
\ket{\phi_{+}} &= \cos \theta \ket{d} + \sin \theta \ket{\vec{0}} \\
\ket{\phi_{-}} &= -\sin \theta \ket{d} + \cos \theta \ket{\vec{0}}
\end{align*}
with the parametrization
\begin{equation*}
\tan (2 \theta) = \frac{2 g}{E_d - E_0}.
\end{equation*}
Starting from the initial state $\ket{d}=\cos\theta\ket{\phi_{+}}-\sin\theta\ket{\phi_{-}}$, the time evolution gives
\begin{align*}
	\langle d|e^{-iHt}|d\rangle&=e^{-iE_+t}\cos^2\theta+e^{-iE_-t}\sin^2\theta\\
	|\langle d|e^{-iHt}|d\rangle|^2&=\cos^4\theta+\sin^4\theta\\
		&\quad+2\cos[(E_+-E_-)t]\cos^2\theta\sin^2\theta.
\end{align*}
The time-dependent term disappears upon time-averaging in Eq.~(\ref{eq:return_probability}) and the return probability is given by the first two terms,
\begin{align*}
Q_d &= \overline{\cos^4 \theta + \sin^4 \theta} = \overline{1-\frac{1/2}{\tan^{-2}(2\theta)+1}} \\
&\approx 1-\frac{1}{2(W+V)}\int_{-(W+V)}^{W+V}dE_0\,\frac{2g^2}{(E_d-E_0)^2+4g^2}.
\end{align*}
At the second line, we have assumed that the energy $E_0$ is uniformly distributed over the interval $[-W-V,W+V]$. Evaluating the integral for $E_d=0$, we find Eq.~(\ref{eq:strong_disorder}).

\section{Calculation of the localization length}
\label{app:localization_length}

\begin{figure}[tb]
\includegraphics[width=\linewidth]{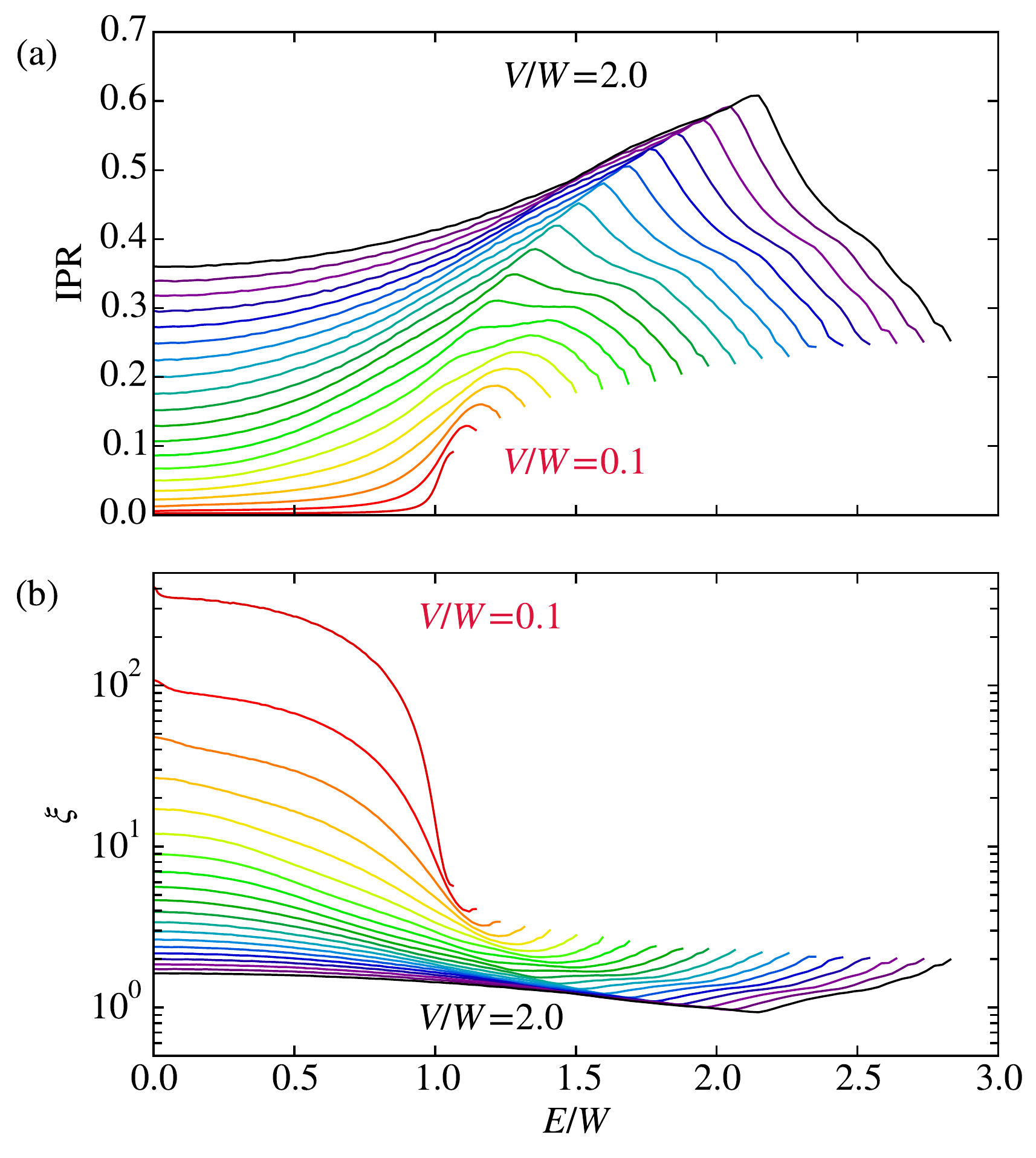}
\caption{(a) Inverse participation ratio $\text{IPR}(E)$ and (b) localization length $\xi(E)$ for a 1D lattice, binned according to eigenenergies and averaged in each bin. The number of bins is 200 and the curves are averages over $N = 2000$ to $N = 10000$ disorder realizations. To achieve convergence of $\xi$, the size of the lattice is increased from $L = 1000$ to $L = 5000$ as disorder gets weaker. The different colors correspond to values of $V$ ranging from $V/W = 0.1$ (red line) to $V/W = 2.0$ (black line) in steps of 0.2. Both $\text{IPR}(E)$ and $\xi$ are symmetric for $E < 0$.}
\label{fig:ipr_xi_1D}
\end{figure}

\begin{figure}[tb]
\includegraphics[width=\linewidth]{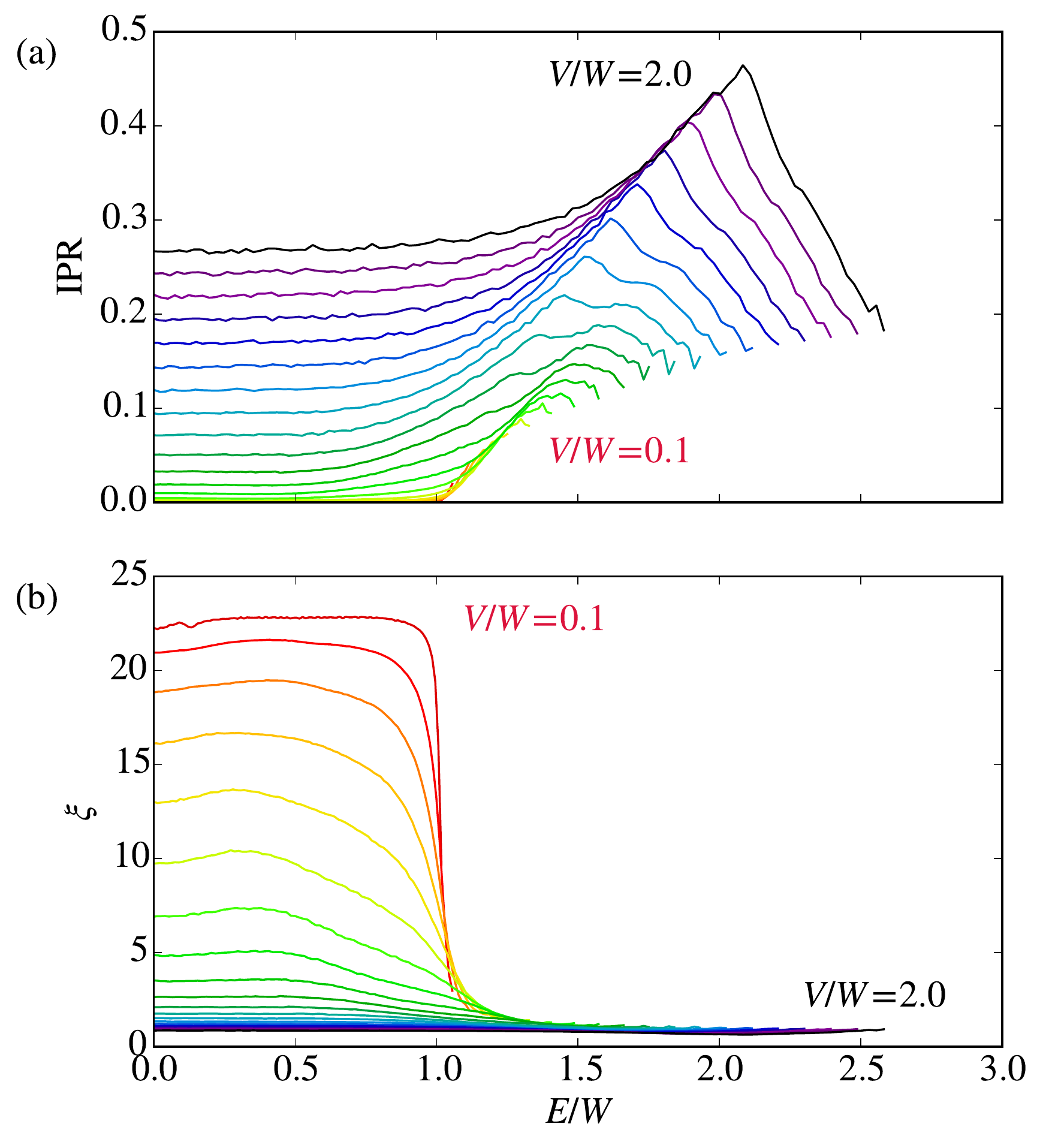}
\caption{IPR and $\xi$ as in Fig.~\ref{fig:ipr_xi_1D} for a 2D lattice of $100 \times 100$ sites. The values of $V$ range from $V/W = 0.1$ to $V/W = 2.0$ in steps of 0.1. For $V/W \lesssim 1.0$, the curves shown here have not yet converged as a function of the lattice size and we perform a finite-size scaling to obtain the values used in the main text. The curves are averages over $N = 200$ disorder realizations.}
\label{fig:ipr_xi_2D}
\end{figure}

\begin{figure}[tb]
\includegraphics[width=\linewidth]{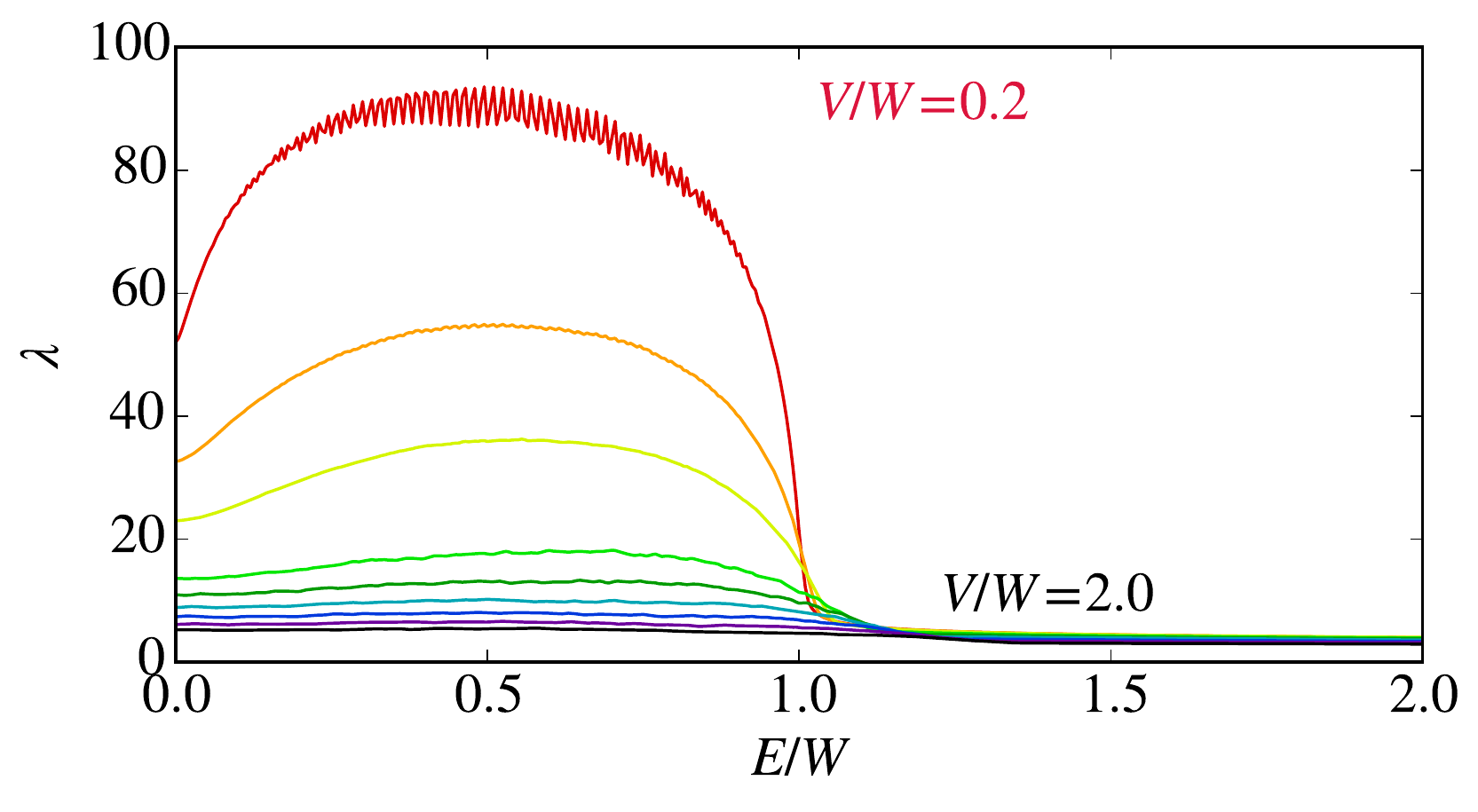}
\caption{Localization length $\lambda$ for a 2D lattice, calculated as in Eq.~(\ref{eq:conductance}) using the Chebyshev expansion. The disordered lattice has $100 \times 100$ sites and is connected with two ideal leads of size $100\times 950$. The Chebyshev expansion order is $M=4000$ and average is made over $N = 10$ to 100 disorder realizations. The oscillations that appear for the weakest disorder $V/W = 0.2$ are due to repeated scattering from the boundaries of the disordered system and the leads. The values of $\lambda$ are larger than $\xi$ in Fig.~\ref{fig:ipr_xi_2D}(b), but the energy dependence is qualitatively similar.}
\label{fig:xi_conductance_2D}
\end{figure}

As discussed in Sec.~\ref{sec:localization_measures}, we calculate the localization length $\xi$ from the inverse participation ratio according to Eq.~(\ref{eq:xi_alpha}). Specifically, we solve the eigenstates by ED, bin the values of $\xi_{\alpha}$ according to the eigenenergies $E_{\alpha}$, and obtain $\xi(E)$ as bin averages which are also averaged over disorder realizations. Figures~\ref{fig:ipr_xi_1D} and \ref{fig:ipr_xi_2D} show $\text{IPR}$ and $\xi$ as functions of energy for a 1D and 2D lattice, respectively. The different colors denote different values of $V$. The curves are averages over 2000 to 10000 realizations of the disorder potential in the case of the 1D lattice and 200 realizations in the case of the 2D lattice. The IPR and the localization length of an eigenstate depend on the energy of the state: states at the band center are less localized than those near the band edges. The IPR does not however grow monotonically with increasing $E$ but has a maximum at a certain energy and then decreases towards the edge of the spectrum $E = W + V$. This decrease is due to rare configurations of the disorder potential where a cluster of neighboring sites has an on-site energy close to $W + V$. Figure~\ref{fig:xi_conductance_2D} shows that the transport localization length calculated with Eq.~(\ref{eq:conductance}) is slightly larger and has a different behavior close to the band edge. A similar difference between the IPR and the Lyapunov exponent, another quantity measuring localization, is discussed in Ref.~\onlinecite{Johri_singular_behavior2012}.

The localization length shown in Figs.~\ref{fig:ipr_xi_2D} and \ref{fig:xi_conductance_2D} is calculated for $L = 100$ and, for the smallest disorder widths $V/W<1$, depends strongly on system size. According to the scaling theory of localization \cite{Abrahams_scaling_theory1979}, $\xi$ scales with system size like $\xi=Lf(\tilde{\xi}/L)$, where the function $f$ is independent of $L$ and $V$ and $\tilde{\xi}$ is the localization length in the thermodynamic limit. Since the sizes reachable by exact diagonalization are limited, we correct the values of $\xi$ using the one-parameter scaling function proposed in Ref.~\onlinecite{Fan_finite-size2014},
\begin{equation}
\frac{\xi}{L} = \frac{1}{k} \ln\left(1 + k\frac{\tilde{\xi}}{L}\right).
\label{eq:scaling_function}
\end{equation}
As our geometry is different from that used in Ref.~\onlinecite{Fan_finite-size2014}, we determine the parameter $k$ by least-squares fitting of the function $\ln[1+k\,\tilde{\xi}(V)/L]/k$ to the values $\xi(L,V)/L$ calculated for $L=30, 40, \dots, 100$, the values $\tilde{\xi}(V)$ being fitting parameters as well. The resulting values of $\tilde{\xi}$ are of the same order of magnitude as those reported in Ref.~\onlinecite{Schreiber_localization_2D_1992}, except for $V/W \lesssim 0.5$ where they are orders of magnitude smaller. The values $\tilde{\xi}$ produced by this procedure are denoted by $\xi$ in the main text, where the values before scaling do not appear. Performing the same finite-size analysis in the case of a 1D lattice does not change the results, indicating that the values of $L$ are sufficiently large for the localization length to be independent of $L$.

\begin{figure}[tb]
\includegraphics[width=0.95\linewidth]{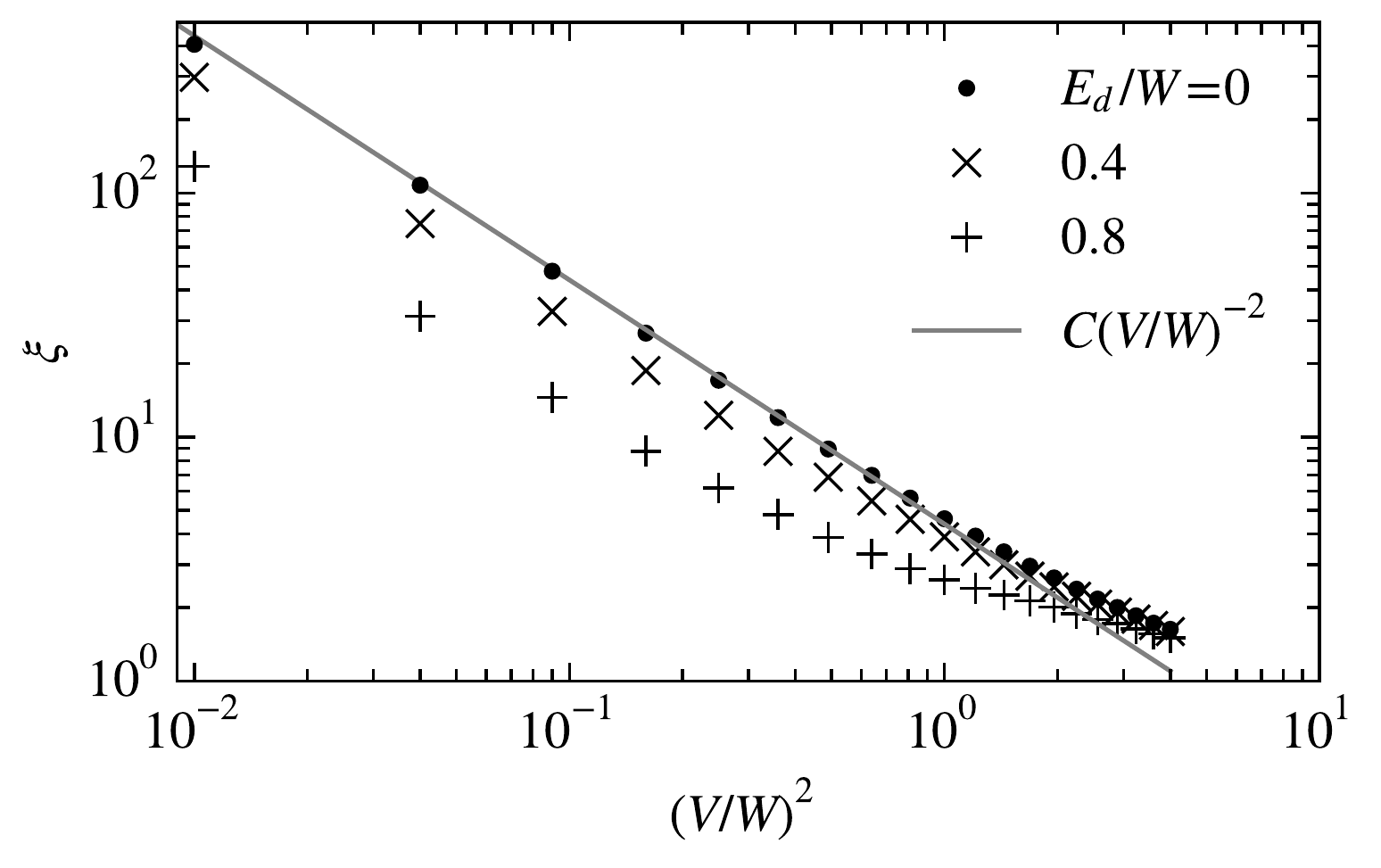}
\caption{Localization length $\xi$ in 1D as a function of $(V/W)^2$ for various energies. The behavior approaches $1/V^2$ at small $V$, showing that $\xi$ is proportional to the mean-free path.}
\label{fig:xi_V_1D}
\end{figure}

\begin{figure}[tb]
\includegraphics[width=0.95\linewidth]{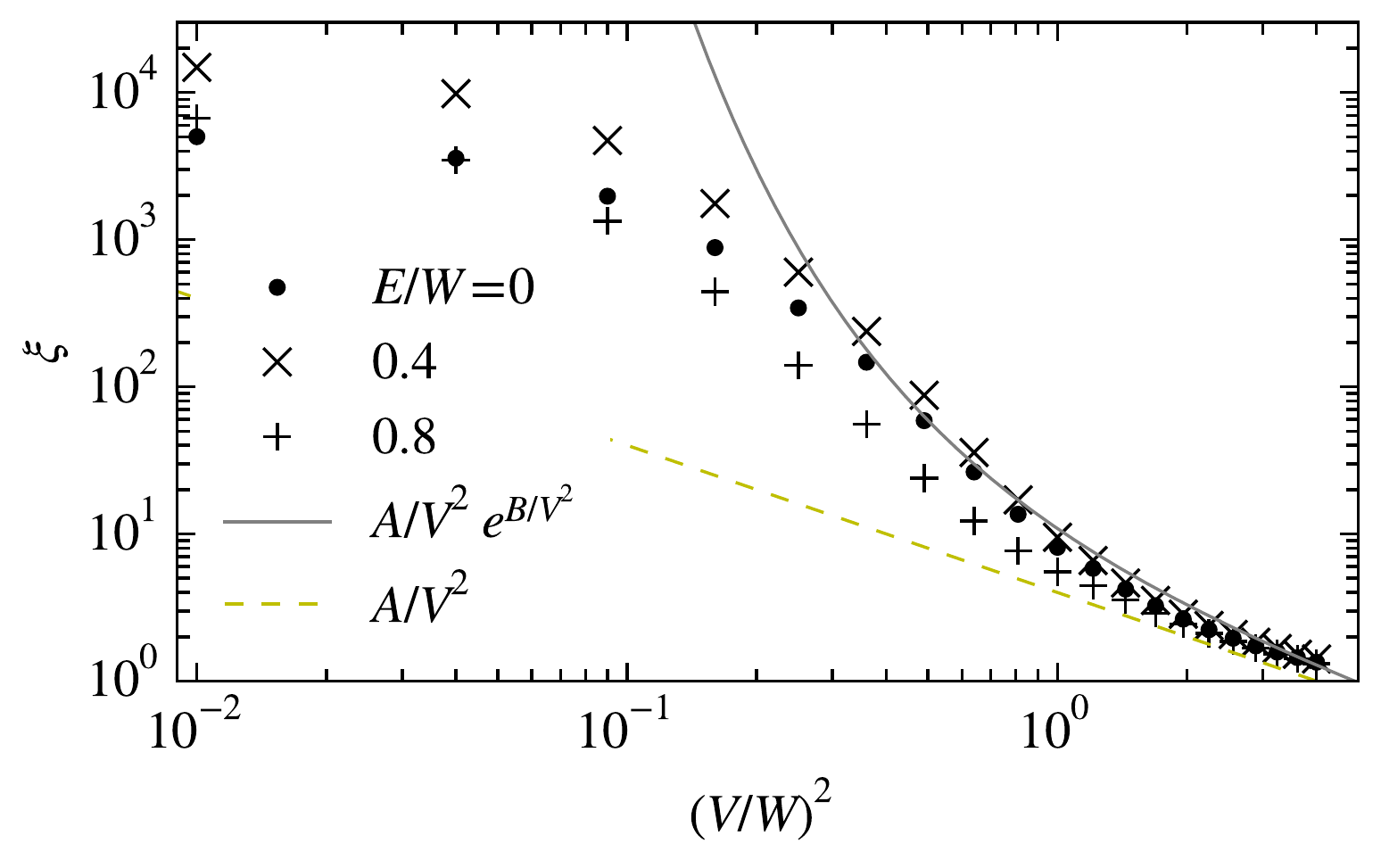}
\caption{Localization length $\tilde{\xi}$ in 2D as a function of $(V/W)^2$ obtained by rescaling the calculated $\xi$ values using Eq.~(\ref{eq:scaling_function}) for various energies. The solid and dashed lines indicate the expected behavior of the localization length and the mean-free path, respectively, at small $V/W$.}
\label{fig:xi_V_2D}
\end{figure}

For weak disorder, the mean-free path varies with disorder strength as $1/V^2$. In 1D, the localization length is expected to be proportional to the mean-free path and therefore also proportional to $1/V^2$, which is confirmed in Fig.~\ref{fig:xi_V_1D} for all the energies shown in the figure. Deviations are noticeable for $V/W>1$. In two dimensions, it is expected that the localization length depends exponentially on the mean-free path. The exponentially large values $\sim e^{1/V^2}$ challenge numerical approaches at small $V$. Our results shown in Fig.~\ref{fig:xi_V_2D} capture the crossover from $1/V^2$ for $V/W\sim1$ to $e^{1/V^2}$ for $V/W<1$, but saturate at small $V$ to values of order $10^4$, showing the limitations of our finite-size scaling approach.

\clearpage
\bibliographystyle{apsrev4-1-with-titles}
\bibliography{bibfile}

\end{document}